\documentclass{elsart}
\usepackage{graphicx}
\usepackage{amssymb}
\usepackage{amsmath}
\begin{document}

\begin{frontmatter}
\title{Steady-state cracking in brittle substrates beneath adherent f\mbox{}ilms: revisited}
\author{E. Matsinos{$^*$}}
\address{Centre for Applied Mathematics and Physics, Zurich University of Applied Sciences, Technikumstrasse 9, P.O. Box, CH-8401 Winterthur, Switzerland}

\begin{abstract}
This is a technical note aiming at the re-examination of the phenomenon of the steady-state cracking in the two-layer system. The method of Suo and 
Hutchinson, as introduced in their $1989$ paper, is followed. Our solution is compared with the one appearing in that paper for the 
substrate-to-f\mbox{}ilm thickness ratio $\lambda_0=10$. We obtain results at three $\lambda_0<10$ values. Combined with the results for $\lambda_0=10$, 
the new sets of values cover thickness ratios between $1$ and $10$, suf\mbox{}f\mbox{}icient for determining crack initiation and propagation in almost 
every relevant problem. We present our results in tables (and f\mbox{}igures), thus facilitating their implementation and use.\\
\noindent {\it PACS:} 62.20.Mk; 81.40.Np 
\end{abstract}
\begin{keyword}
Steady-state cracking; two-layer system
\end{keyword}
{$^*$}{E-mail address: evangelos.matsinos@zhaw.ch}
\end{frontmatter}

\section{\label{sec:Introduction}Introduction}

The phenomenon of the steady-state cracking in the two-layer system was investigated in the work of Suo and Hutchinson \cite{SH} more than thirty years 
ago. In the present study, we will assume familiarity with the content of that paper; def\mbox{}initions for some of the quantities, which were introduced 
therein, will be given for reasons of clarity. Unless mentioned otherwise, the Suo-Hutchinson notation (of that paper) will be closely followed.

A two-layer system is produced when a block of material $1$, typically a thin metallic plate or f\mbox{}ilm, is attached to (e.g., glued or deposited upon) 
a block of material $2$. In view of the interest in terms of application, we will assume that material $2$ is a semiconductor; by no means, should this be 
taken as a restriction of the method. Under the inf\mbox{}luence of external factors (e.g., via the application of mechanical force), a crack may appear 
and propagate at some depth in material $2$. Steady-state cracking implies the propagation of the crack parallel to the interface of the two materials; this 
is one of the possible outcomes and, perhaps, the most interesting one in practical applications. Other possibilities include substrate cracking in the direction 
perpendicular to the interface, interface debonding, channelling, f\mbox{}ilm buckling, etc. For details, the reader is addressed to Ref.~\cite{HS}.

The interest in the phenomenon of cracking is attributable to the production of thin slices of semiconductors (e.g., silicon); typical 
thicknesses span tens to a few hundreds of microns. At present, this technology of\mbox{}fers the only available solution for large-scale 
production in case that the desirable thickness of the extracted slice exceeds a few tens of microns.

Despite that the theoretical background in Ref.~\cite{SH} is independent of the specif\mbox{}ic technique used in the cracking, it was 
nonetheless developed for stresses which are mechanically induced, i.e., for crack initiation and propagation under the application 
of mechanical force on the sides of the two-layer system (e.g., see Fig.~2 of Ref.~\cite{SH}). Another technique has gained 
ground in the recent years, namely, that of thermally-induced stress (e.g., see Refs.~\cite{HS} and \cite{Dross}), featuring the 
insertion of the two-layer system into an environment of temperature (far) below the one corresponding to the preparation (paste/deposition 
of material $1$ on material $2$). Given the dif\mbox{}ference of the coef\mbox{}f\mbox{}icients of thermal expansion of the metal and of the semiconductor, 
the f\mbox{}ilm bends f\mbox{}irst, thus inducing stress onto the substrate, which (depending on the conditions, materials, and thicknesses) may 
lose its cohesion and break. Finally, the metallic f\mbox{}ilm tears of\mbox{}f a layer of the semiconductor (which may be subsequently retrieved 
with further processing, e.g., via the use of a metal-etching solution \cite{Dross}).

Irrespective of the technique employed in the cracking (i.e., mechanically- or thermally-induced stress, or a combination of the two), 
the basic phenomenon may be described by a simple model of ef\mbox{}fective longitudinal (mechanical) loads and moments \cite{SH}, which 
are linked to the stress intensity factors $K_I$ and $K_{II}$ in the volume of material $2$ (see Eq.~(6) of Ref.~\cite{SH}). Suo 
and Hutchinson introduced a parameterisation (through their Eqs.~(8)-(11)) in order to enable the evaluation of $K_I$ and $K_{II}$ 
in material $2$ from the ef\mbox{}fective mechanical load and moment; apart from the physical constants of the materials, introduced 
by their Eq.~(9) is the angle $\omega$, the only unknown in the evaluation. Having obtained $\omega$ as a function of the depth in 
material $2$ (for the materials being used), one may determine $K_I$ and $K_{II}$ in material $2$ and decide if (and where) a crack 
will be initiated and how it is expected to propagate. Evidently, crack initiation will occur at a position where the total stress exceeds 
the fracture toughness of the substrate $K_{Ic}$ (which, in silicon, is about $0.9$MPa$\sqrt{\rm m}$). In case that $K_{II}=0$ in a path 
parallel to the interface of the two materials, the crack is expected to travel along that path \cite{SH}. By varying the external conditions 
(i.e., the mechanical forces and/or the temperature dif\mbox{}ference), material $1$, its thickness, as well as the thickness of the silicon substrate, 
one may control the thickness of the peeled-of\mbox{}f semiconductor slice (and the quality of the extraction, ref\mbox{}lected in the variation of the 
thickness over the slice).

At this point, the reader might wonder why should the subject be revisited. There are a number of reasons calling for further results. The 
f\mbox{}irst and most important reason concerns the extraction of $\omega$ for more values of the substrate-to-f\mbox{}ilm thickness ratio 
(hereafter denoted as $\lambda_0$); the two cases, which are fully documented in Ref.~\cite{SH}, i.e., $\lambda_0^{-1}=0$ and $0.10$, are hardly 
suf\mbox{}f\mbox{}icient when determining the stress intensity factors in the general case (i.e., at arbitrary $\lambda_0^{-1}$). (It must 
be noted that the $\omega$ values of \cite{SH} for the $\lambda_0^{-1}=0.05$ case are not useful, as the corresponding corrections for 
$\beta \neq 0$ have not been reported.) In the present paper, we plan to extract the $\omega$ values for $\lambda_0^{-1}=0.25$, $0.50$, and 
$1.00$, i.e., for smaller substrate-to-f\mbox{}ilm thickness ratios than those treated in Ref.~\cite{SH}. We also 
plan to extract the $\omega$ values for $\lambda_0^{-1}=0.10$ and compare our results to those of Ref.~\cite{SH}.

Concerning the determination of the $\omega$ values at arbitrary $\lambda_0^{-1} \neq 0$ or $0.10$, Suo and Hutchinson \cite{SH} wrote: 
`It is believed that $\omega$ is analytic in $\lambda_0^{-1}$ for large $\lambda_0$, and therefore $\omega$ can be well approximated 
by linear interpolation in $\lambda_0^{-1}$ between $\lambda_0^{-1}=0$ and $0.10$, where values of $\omega$ have been tabulated.' Upon 
inspection, however, of the entries of their Tables $3$-$5$, one cannot but feel somewhat uneasy about that statement. Were it valid, 
one should be able to obtain the solution for $\lambda_0^{-1}=0.05$ (listed in their Table $4$) as average values of the corresponding 
entries of their Tables $3$ and $5$. As a matter of fact, the solution of their Table $4$ \emph{systematically} exceeds this average 
by anything up to $1.6^0$, the average dif\mbox{}ference being equal to $0.66^0$. Additionally, there are large curvature ef\mbox{}fects 
for $\lambda \gtrsim 3$, the worst of which show up at $\alpha=0$ and $\lambda=6$, where the three solutions are given as: $52.0$, 
$51.0$, and $46.8^0$ for $\lambda_0^{-1}=0$, $0.05$, and $0.10$, respectively. For $9$ out of the $54$ cases in Tables $3$-$5$ (for 
which a comparison is possible), monotony does not seem to hold at all; in this respect, the $\omega$ values at $\alpha=0.8$ are 
particularly problematic.

Regarding the corrections which must be applied (to the $\omega$ values) because of a non-zero $\beta$, some numbers have appeared 
in Tables $1$ and $2$ of Ref.~\cite{SH}, yet in a form which makes the application of the correction dif\mbox{}f\mbox{}icult. The improvement we 
propose at this point is simple. We found that, for a given ($\alpha$, $\lambda_0$, $\lambda$) combination, the assumption that 
the correction (def\mbox{}ined as the $\omega$ value at $\beta \neq 0$ minus the one at $\beta=0$) scales with $\beta$ is a good approximation. 
We will therefore give these multiplicative factors for each ($\alpha$, $\lambda_0$, $\lambda$) combination treated; the corrections 
may be easily obtained via a simple multiplication with the $\beta$ value. As the corrections for negative and positive 
$\beta$ values came out dif\mbox{}ferent, two sets of correction factors for each ($\alpha$, $\lambda_0$, $\lambda$) combination will be given.

Last but not least, the present work serves one additional purpose. A detailed technical report on how to obtain the $\omega$ values 
will be surely useful in the future to anyone who has interest in studying further the phenomenon of cracking in the two-layer (perhaps, 
also in the three-layer) system. We f\mbox{}ind it surprising that, despite the fairly-broad use of the results of Ref.~\cite{SH} (e.g., in 
f\mbox{}inite-element calculations in Materials Science), no ef\mbox{}fort has been made in more than thirty years (to the best of the author's 
knowledge) towards a further study of multi-layer systems within the framework set forth in that pioneering work.

\section{\label{sec:Method}Method}

\subsection{\label{sec:Model}The Suo-Hutchinson model}

The thickness of material $1$ is denoted by $h$; all relevant lengths in the problem will be expressed as multiples of $h$. The 
thickness of material $2$ is denoted as $\lambda_0 h$. We will generally use the inverse of $\lambda_0$ in this paper; in this 
case, the solution for an inf\mbox{}initely-thick substrate corresponds to $\lambda_0^{-1}=0$, for a substrate ten times 
thicker than the f\mbox{}ilm to $\lambda_0^{-1}=0.10$, and so on.

\begin{figure}
\begin{center}
\includegraphics [angle=270,width=15.5cm] {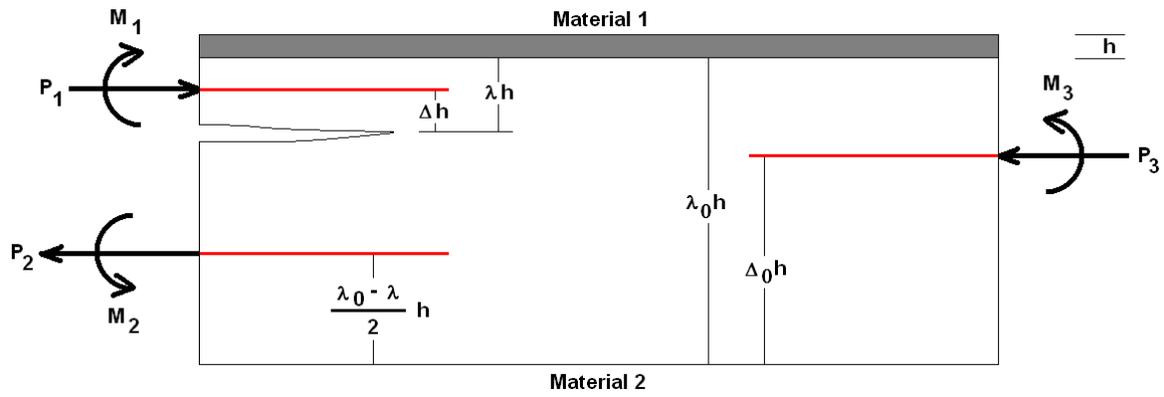}
%\vspace{-6cm}
\caption{\label{fig:Model}The Suo-Hutchinson model. The red lines indicate the positions of the neutral axes.}
\end{center}
\end{figure}

Shown in Fig.~\ref{fig:Model}, is a crack propagating at depth $\lambda h$ inside the material $2$ (the depth is equal to $0$ at 
the interface of materials $1$ and $2$). Let us assume that the crack is propagating under the exertion of the longitudinal loads $P_i$ and moments 
$M_i, i=1,2,3$ (both expressed `per unit of thickness'), applied to the two-layer system at the positions corresponding to the 
neutral axes. The positions of the neutral axes are obtained via the formulae
\begin{equation} \label{eq:DefinitionOfDelta}
\Delta = \frac{\lambda^2 + 2 \Sigma \lambda + \Sigma}{2 (\lambda + \Sigma)}
\end{equation}
and
\begin{equation} \label{eq:DefinitionOfDelta0}
\Delta_0 = \frac{\lambda_0^2 + 2 \Sigma \lambda_0 + \Sigma}{2 (\lambda_0 + \Sigma)} \, ,
\end{equation}
where the quantity $\Sigma$ is the stif\mbox{}fness ratio of materials $1$ and $2$
\begin{equation} \label{eq:DefinitionOfSigma}
\Sigma = \frac{1+\alpha}{1-\alpha} \, ,
\end{equation}
with
\begin{equation} \label{eq:DefinitionOfAlpha}
\alpha = \frac{\Gamma (\kappa_2 + 1) - (\kappa_1 +1)}{\Gamma (\kappa_2 + 1) + (\kappa_1 +1)} \, ;
\end{equation}
$\Gamma=\frac{\mu_1}{\mu_2}$ denotes the ratio of the shear moduli $\mu$ (the shear modulus is the ratio of the shear stress to the shear 
strain) of the two materials; $\kappa = 3-4\nu$ for plane strain and $(3-\nu)/(1+\nu)$ for 
plane stress; f\mbox{}inally, $\nu$ denotes Poisson's ratio (the transverse-to-axial strain ratio when the material is stretched). The quantity 
$\alpha$ is one of the two parameters introduced by Dundurs \cite{Dundurs} for the description of the mechanical properties of bimaterial 
systems; the second parameter is another combination of $\Gamma$ and of the $\kappa_i$'s
\begin{equation} \label{eq:DefinitionOfBeta}
\beta = \frac{\Gamma (\kappa_2 - 1) - (\kappa_1 - 1)}{\Gamma (\kappa_2 + 1) + (\kappa_1 +1)} \, .
\end{equation}
It has been argued in the literature that $\beta$ ref\mbox{}lects the oscillatory behaviour of the crack tip as the crack propagates in material $2$. 
If materials $1$ and $2$ are equally stif\mbox{}f, then $\Sigma=1$, in which case 
Eqs.~(\ref{eq:DefinitionOfDelta},\ref{eq:DefinitionOfDelta0}) reduce to $\Delta = (\lambda + 1) /2$ and $\Delta_0 = (\lambda_0 + 1) /2$; therefore, 
in the trivial case of identical materials $1$ and $2$, the neutral axes correspond, as expected, to the midpoints of the left (above the crack 
level) and right sides.

Suo and Hutchinson show (see Appendix A of Ref.\cite{SH}) that the phenomenon is adequately described on the basis of one ef\mbox{}fective longitudinal 
load $P$ and one moment $M$. They come to this simplif\mbox{}ication after considering the two conditions for no net translation and rotation of the 
two-layer system (i.e., their Eqs.~(3)), as well as the stress inside material $1$ before and after the crack tip; the ef\mbox{}fective quantities 
$P$ and $M$ may be obtained via the relations
\begin{equation} \label{eq:DefinitionOfP}
P \equiv P_1 - C_1 P_3 - C_2 M_3 / h
\end{equation}
and
\begin{equation} \label{eq:DefinitionOfM}
M \equiv M_1 - C_3 M_3 \, ,
\end{equation}
with
\begin{equation} \label{eq:DefinitionOfC1}
C_1 = A / A_0 \, ,
\end{equation}
\begin{equation} \label{eq:DefinitionOfC2}
C_2 = A [ (\lambda_0 -\Delta_0) - (\lambda - \Delta)] / I_0 \, ,
\end{equation}
and
\begin{equation} \label{eq:DefinitionOfC3}
C_3 = I / I_0 \, .
\end{equation}
The expressions for the ef\mbox{}fective cross section $A$ and the moment of inertia (per unit length) $I$ may be found in Appendix 
A of that paper:
\begin{equation} \label{eq:DefinitionOfA}
A = \lambda + \Sigma
\end{equation}
and 
\begin{equation} \label{eq:DefinitionOfI}
I = \{ \Sigma [3 (\Delta - \lambda)^2 - 3 (\Delta - \lambda) + 1] + 3 \Delta \lambda (\Delta - \lambda) + \lambda^3 \} /3 \, .
\end{equation}
Evidently,
\begin{equation} \label{eq:DefinitionOfA0}
A_0 = \lambda_0 + \Sigma
\end{equation}
and 
\begin{equation} \label{eq:DefinitionOfI0}
I_0 = \{ \Sigma [3 (\Delta_0 - \lambda_0)^2 - 3 (\Delta_0 - \lambda_0) + 1] + 3 \Delta_0 \lambda_0 (\Delta_0 - \lambda_0) + \lambda_0^3 \} /3 \, .
\end{equation}

Suo and Hutchinson subsequently proceed to associate the ef\mbox{}fective quantities $P$ and $M$ directly with the stress intensity factors $K_I$ and 
$K_{II}$ in the interior of material $2$, proposing a relation (their Eq.~(6)) reading as
\begin{equation} \label{eq:DefinitionOfStresses}
K_I + i K_{II} = \frac{1}{\sqrt{2}} \Big( a \frac{P}{\sqrt{Uh}} + b \frac{M}{\sqrt{Vh^3}} \Big) \, ,
\end{equation}
where $a$ and $b$, dimensionless complex quantities of unit modulus, depend on $\alpha$, $\beta$, $\lambda_0$, and $\lambda$; the functions 
$U$ and $V$ have been given in their Appendix A.
\begin{equation} \label{eq:DefinitionOfU}
U^{-1} = A^{-1} + \frac{1}{\lambda_0 - \lambda} + \frac{12 [\Delta + (\lambda_0 - \lambda)/2]^2}{(\lambda_0 - \lambda)^3}
\end{equation}
and 
\begin{equation} \label{eq:DefinitionOfV}
V^{-1} = I^{-1} + \frac{12}{(\lambda_0 - \lambda)^3} \, .
\end{equation}
The formalism is completed after putting $a$ and $b$ in the forms
\begin{equation} \label{eq:DefinitionOfa}
a = e^{i \omega}
\end{equation}
and
\begin{equation} \label{eq:DefinitionOfb}
b = - i e^{i (\omega + \gamma)} \, ,
\end{equation}
where the angle $\gamma$ may be obtained (according to the third of their equations appearing under (A$8$)) by using the formula
\begin{equation} \label{eq:DefinitionOfGamma}
\gamma = \sin^{-1} \Big\{ \frac{12 \sqrt{UV} [\Delta + (\lambda_0 - \lambda)/2]}{(\lambda_0 - \lambda)^3} \Big\} \, .
\end{equation}

Obviously, the only quantity which must be known in order that the stress intensity factors $K_I$ and $K_{II}$ be determined is the angle 
$\omega$, which should be considered a function of $\alpha$, $\beta$, $\lambda_0$, and $\lambda$. After having set the $\lambda$-dependence
of $\omega$ for the specif\mbox{}ic ($\alpha$, $\beta$, $\lambda_0$) combination, one may directly obtain the stress intensity factors in the entire 
volume of the semiconductor, and assess if (and where) a crack will be initiated and how it is expected to propagate. We will deal with the 
determination of the $\omega$ values in the following section.

\subsection{\label{sec:Determination}Determination of the $\omega$ values}

The extraction of the angle $\omega$ for each ($\alpha$, $\beta$, $\lambda_0$, $\lambda$) combination is a rather involved issue; it is recommended 
that the interested reader consult Appendices B and C of Ref.~\cite{SH} for the details. Herein, we intend only to draw attention to a number of relevant 
issues. The integral equation, which is to be solved (i.e., Eq.~(B$11$) of \cite{SH}), reads as
\begin{equation} \label{eq:B11}
\int_{-1}^{1} \frac{\bar{A}(t)}{u-t} dt + \int_{-1}^{1} \frac{F_1(\zeta) A(t) + [1+t+F_2(\zeta)] \bar{A}(t)}{(1+t)^2} dt = 0 \, ,
\end{equation}
where $\bar{A}(t)$ denotes the complex conjugate of $A(t)$. The quantities $u$ and $t$ are real numbers satisfying $-1<u,t<1$; 
$\zeta=\frac{2(u-t)}{(u+1)(t+1)}$. The complex functions $F_{1,2}(t)$ are def\mbox{}ined in Appendix C of Ref.~\cite{SH}. The 
complex function $A(t)$ is expanded in terms of the Chebyshev polynomials of the f\mbox{}irst kind $T_{k} (t)$:
\begin{equation} \label{eq:B12}
A(t) = \Big( \frac{1-t}{2} \Big) ^{-1/2} \big[ B(-\infty)+(1+t) \sum_{k=1}^{N} a_k T_{k-1} (t) \big] \, ,
\end{equation}
where $B(-\infty)$ is a complex constant with known imaginary part (see the f\mbox{}irst of Eqs.~(B$7$) of Ref.\cite{SH}) and $a_k$ are complex coef\mbox{}f\mbox{}icients, 
the real and imaginary parts of which must be determined. Without loss of generality, the real part of $a_N$ may be set to $0$, after which there 
are $2N$ unknowns in the problem (i.e., $2N-1$ numbers associated with the coef\mbox{}f\mbox{}icients $a_k$ and $\Re [B(-\infty)]$). The integral equation (\ref{eq:B11}) 
is solved on $N$ $u_i$ points, chosen to be the $N$ Gauss-Legendre points in $[-1,1]$, and the decomposition of the set into real and imaginary parts 
yields $2N$ equations, i.e., as many as unknowns in the problem. The set of equations may be written in a compact form as
\begin{equation} \label{eq:MatrixEquation}
\sum_{k=1}^{N} [ a_k I_1(u,k) + \bar{a}_k I_2 (u,k)] + \Re [B(-\infty)] I_3 (u) = I_4 (u) \, .
\end{equation}
As only $I_1 (u,k)$ has been explicitly given in Ref.~\cite{SH}, we will now list all expressions;
\begin{equation} \label{eq:I1}
I_1 (u,k) = \int_{-1}^{1} \frac{F_1(\zeta)}{1+t} \Big( \frac{1-t}{2} \Big)^{-1/2} T_{k-1} (t) dt \, ,
\end{equation}
\begin{align} \label{eq:I2}
I_2 (u,k) = & \int_{-1}^{1} \frac{1}{u-t} \Big( \frac{1-t}{2} \Big)^{-1/2} (1+t) T_{k-1} (t) dt + \nonumber \\
& \int_{-1}^{1} \frac{1+t+F_2(\zeta)}{1+t} \Big( \frac{1-t}{2} \Big)^{-1/2} T_{k-1} (t) dt \, ,
\end{align}
\begin{align} \label{eq:I3}
I_3 (u) = & \int_{-1}^{1} \frac{1}{u-t} \Big( \frac{1-t}{2} \Big)^{-1/2} dt + \int_{-1}^{1} \frac{F_1(\zeta)}{(1+t)^2} \Big( \frac{1-t}{2} \Big)^{-1/2} dt + \nonumber \\
& \int_{-1}^{1} \frac{1+t+F_2(\zeta)}{(1+t)^2} \Big( \frac{1-t}{2} \Big)^{-1/2} dt \, ,
\end{align}
and
\begin{align} \label{eq:I4}
I_4 (u) = - i \Im [B(-\infty)] \Big\{ - & \int_{-1}^{1} \frac{1}{u-t} \Big( \frac{1-t}{2} \Big)^{-1/2} dt + \nonumber \\ 
& \int_{-1}^{1} \frac{F_1(\zeta)}{(1+t)^2} \Big( \frac{1-t}{2} \Big)^{-1/2} dt - \nonumber \\ 
& \int_{-1}^{1} \frac{1+t+F_2(\zeta)}{(1+t)^2} \Big( \frac{1-t}{2} \Big)^{-1/2} dt \Big\} \, .
\end{align}
Given the asymptotic behaviour of the functions $F_{1,2}(t)$ (see Eqs.~(B$3$) of \cite{SH}), all integrands behave well as $t \to -1$ (equivalently, 
$\zeta \to \infty$) and all integrals exist. The quantities $I_{2,3,4}$ contain terms in which Cauchy principal values must be determined (i.e., 
the integrands involving the inverse of $u-t$). The method of Longman \cite{Longman} has been used to estimate the contribution in $I_2 (u,k)$; the 
value of the integral, appearing in $I_{3,4}$, may be obtained analytically.

We now touch upon one important point, namely the choice of the number $N$ of Chebyshev polynomials to be used in the expansion of $A (t)$. In 
Ref.\cite{SH}, it is noted that the results had been obtained `with $N$ between $10$ and $15$.' We have varied $N$ between $6$ and $16$ (with a 
step of $2$), but we have not been able to observe true convergence in the extracted $\omega$ values; the $\omega$ values (in most cases) come out 
close (within a range of $2^0$), yet the characteristic trend, resembling an approach to an asymptotic value, was rarely 
observed. Concerning the point of whether true convergence had been observed in the cases reported in Ref.~\cite{SH}, the text is of little help. 
Suo and Hutchinson write at the end of their Appendix B: `The consistency check was satisf\mbox{}ied to better than $0.1 \%$. It is believed that the 
accuracy in $\omega$ is comparable.' One may speculate on the meaning of these statements. If by `consistency check' the authors imply the 
fulf\mbox{}illment of the convergence criterion (i.e., the $\omega$ values for successive $N$ runs dif\mbox{}fer less than $0.1 \%$ of their average), the fact 
that two successive $\omega$ values happen to come out `close enough' should not be taken as evidence of true convergence; for instance, we 
observed (several times) that the $\omega$ values for $N=10$ and $12$ came out close, but at least one of the subsequent values (i.e., those 
corresponding to $N=14$ or $16$) was distant, yet not distant enough to be considered an outlier. Although Suo and Hutchinson might have imposed 
additional constraints (e.g., on the behaviour of the coef\mbox{}f\mbox{}icients $a_k$), which have not been mentioned in their paper, it seems dif\mbox{}f\mbox{}icult to 
obtain a solution in this problem, accurate down to the $0.1 \%$ level. We have not been able to resolve this issue with the authors of 
Ref.~\cite{SH}. Lacking the exact details on how the results of Ref.~\cite{SH} were obtained at this point (and whether true convergence had been 
observed in the cases reported), we will follow another strategy in determining the f\mbox{}inal $\omega$ value (and its associated uncertainty) in each 
($\alpha$, $\beta$, $\lambda_0$, $\lambda$) combination; this is the subject of the next section.

After the solution for the coef\mbox{}f\mbox{}icients $a_k$ and for $\Re [B(-\infty)]$ is obtained, the $\omega$ value is extracted using the relation
\begin{equation} \label{eq:Omega}
\omega = - \tan^{-1} \Big\{ \frac{\Im[B(-\infty) + 2 \sum_{k=1}^{N} a_k ]}{\Re[B(-\infty) + 2 \sum_{k=1}^{N} a_k ]} \Big\} - \gamma \, ,
\end{equation}
the angle $\gamma$ being taken from Eq.~(\ref{eq:DefinitionOfGamma}).

\subsection{\label{sec:Technical}Some technical issues}

\subsubsection{\label{sec:Extraction}Extraction of the $\omega$ values}

As six $N$ values have been used in Eq.~(\ref{eq:B12}), six $\omega$ values are extracted in each ($\alpha$, $\beta$, $\lambda_0$, $\lambda$) 
combination. In a perfect world, it is expected that these values show trends of convergence, i.e., of `approaching' the asymptotic value, 
def\mbox{}ined as the limit when $N \to \infty$. As mentioned earlier, we have not been able to observe traces of such a behaviour in the majority of the 
cases examined. Presumably, this failure is due to the combination of two ef\mbox{}fects: a) imprecision in the evaluation of the integrals entering 
the terms $I_i$ and b) `noise' introduced by the inversion of large matrices for the extraction of the unknowns $a_k$ and $\Re [B(-\infty)]$ 
(when $N=16$, a $32 \times 32$ matrix must be inverted).

In the absence of signs of convergence in the series of the extracted values of $\omega$, we have decided to determine one average $\omega$ 
value, as well as the associated uncertainty per case, i.e., per ($\alpha$, $\beta$, $\lambda_0$, $\lambda$) combination, and further process 
the resulting data. It is not easy, however, to produce meaningful averages and uncertainties from only six data points, especially so in the 
presence of outliers. Consequently, the f\mbox{}irst step in the evaluation of $\omega$ must involve the employment of an ef\mbox{}f\mbox{}icient 
algorithm for outlier rejection. A few outlier-detection algorithms have been tested. The f\mbox{}irst successful results were obtained with Grubbs's 
outlier test \cite{Grubbs,Stefansky}; although the test checks the input data for the presence of one single outlier, it may enable, if iterated, 
the exclusion of several deviant points. In a number of cases, however, the process of applying Grubbs's test iteratively terminated at the f\mbox{}irst 
step, declaring no outliers, though the visual inspection of the data easily (and clearly) identif\mbox{}ied two outliers. (This is a known problem for 
Grubbs's outlier test.) As a result, we decided to apply a more recent algorithm, namely, Rosner's generalised ESD (Extreme Studentised Deviate) 
test \cite{Rosner}. In this algorithm, a set of $N_k$ data points is tested for the presence of exactly $1$, $2$, \dots, $N_a$ outliers, where 
$N_a$ is a user-def\mbox{}ined integer satisfying the condition $N_a < N_k$ (in fact, it does not make much sense to perform the test with $N_a$ 
exceeding $N_k/2$); the advantage of Rosner's test is that the `optimal' number of outliers is \emph{extracted} from the input data set.

The maximal number of outliers $N_a$ was set to $3$ in the analysis. Three signif\mbox{}icance levels were used: $0.20$, $0.10$, and $0.05$; large 
values of the signif\mbox{}icance level result in the detection of many outliers, small ones leave outliers in the data. The results for the three 
signif\mbox{}icance levels were always compared prior to making decisions. At each signif\mbox{}icance level, Rosner's test is (in our case) performed 
for exactly one, two, and three outliers in each input data set. At each of these three steps, the set of candidate outliers is accepted or 
rejected on the basis of the comparison of the score value of the test statistic with a critical value, which depends on the number of `good' points 
(those points which do not belong to the set of candidate outliers) and on the signif\mbox{}icance level. Of course, if no score value exceeds the 
critical one at any step, the input data set contains no outliers.

\begin{figure}
\begin{center}
\includegraphics [angle=270,width=15.5cm] {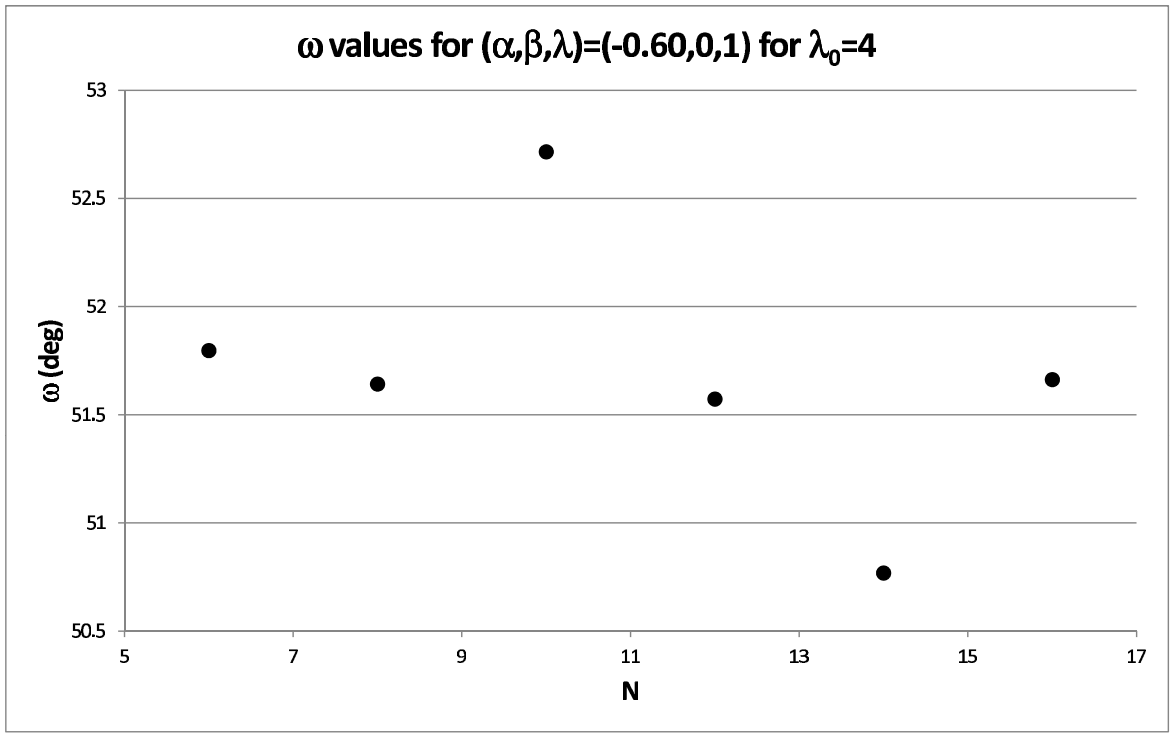}
%\vspace{-6cm}
\caption{\label{fig:Omega}The $\omega$ values extracted for ($\alpha$, $\beta$, $\lambda_0^{-1}$, $\lambda$) $=$ ($-0.60$, $0$, $0.25$, $1.0$); $N$ 
denotes the number of Chebyshev polynomials of the f\mbox{}irst kind used in the expansion of $A(t)$ according to Eq.~(\ref{eq:B12}).}
\end{center}
\end{figure}

One example of extracted $\omega$ values, obtained for ($\alpha$, $\beta$, $\lambda_0^{-1}$, $\lambda$) $=$ ($-0.60$, $0$, $0.25$, $1.0$), 
is shown in Fig.~\ref{fig:Omega}. We will now examine what actually happens if the $\omega$ values, shown in this f\mbox{}igure, are submitted to Rosner's 
test. Let us f\mbox{}ix the conf\mbox{}idence level to $0.20$. The assumption that the data contains exactly one outlier is f\mbox{}irst tested. The highest $\omega$ 
value in the sample (which is the most deviant data point) is tested f\mbox{}irst; its score value is equal to $1.6462$, whereas the critical score value is (for the 
conf\mbox{}idence level used and 
for a six-element set) $1.7289$. As a result, the point may not be considered an outlier (at this step). (Grubbs's test would have terminated at 
this point.) The sample is then tested for the presence of exactly two outliers (the two candidates now being the previous candidate outlier 
and the most distant value after the f\mbox{}irst candidate outlier is removed from the data, i.e., the lowest $\omega$ value). The score value for 
the second potential outlier is equal to $1.7535$, whereas the critical score value (for the conf\mbox{}idence level used and for a f\mbox{}ive-element set) 
is $1.6016$. Therefore, the result of the test for two outliers is af\mbox{}f\mbox{}irmative, the largest and smallest values in the data set having been 
established as true outliers. The sample is then tested for exactly three outliers (the two already-established outliers and the most deviant 
of the remaining values); tested at this step is the value obtained for $N=6$. The score value for the third potential outlier is equal to 
$1.3680$, whereas the critical score value (for the conf\mbox{}idence level used and for a four-element set) is $1.4250$. The f\mbox{}inal conclusion is that 
there are only two outliers in the data (the highest number of established outliers in the $N_a$ steps of the test). The results of this test 
for the other two signif\mbox{}icance levels agree with the presence of only two outliers in the data (identical to those established at the $0.20$ 
signif\mbox{}icance level). The two outliers were removed from the input data and the average value and the associated uncertainty (the standard error 
of the means) were estimated from the remaining measurements and stored as f\mbox{}inal result of the processing for the specif\mbox{}ic ($\alpha$, $\beta$, 
$\lambda_0$, $\lambda$) combination.

Described in the example of Fig.~\ref{fig:Omega} is one clear-cut case, as the results obtained at the three signif\mbox{}icance levels agree. In many cases, 
however, the numbers of outliers for the three signif\mbox{}icance levels come out dif\mbox{}ferent. In order to avoid glitches in the application of the algorithm, 
we decided to set up an interactive procedure and also visually inspect the data, parallel to the application of the algorithm. Special attention was paid when the 
$\omega$ values came out close (e.g., not more than a few tenths of one degree apart); in that case, the test was 
not performed and all data were accepted in the evaluation of the $\omega$ average and the accompanying uncertainty. Given the smallness of the initial 
sample, potentially problematic are the cases with one-sided (i.e., lying on the same side of the average) outliers; one may easily obtain an erroneous 
average. There is less risk in case that the two outliers lie on either side of the average (only the uncertainty depends on the treatment, i.e., on 
the exclusion or inclusion of both values); fortunately, most cases with two outliers fall in this last category.

The $\omega$ values at $\lambda_0 \leq 4$ contain very few outliers. On the contrary, the $\lambda_0=10$ data contain many values 
which seem to be `out of place', especially so in the small-$\lambda$ region; additionally, the $N=6$ and $16$ solutions contain values which f\mbox{}luctuate to the 
extent that the rejection of the corresponding sets is called for more often than not. To overcome the problem of submitting obvious outliers to the outlier-detection 
algorithm, we have decided to f\mbox{}irst visually inspect the tables of the data obtained at f\mbox{}ixed values of $\alpha$, $\beta$, and 
$\lambda_0$; in these tables of $\omega$ values, the rows indicate dif\mbox{}ferent $\lambda$ values, whereas the columns dif\mbox{}ferent $N$ values. After the 
removal of the obvious outliers in each table, there are three options one may follow.
\begin{itemize}
\item Perform an overall f\mbox{}it to all surviving data (assuming constant uncertainties) and present the f\mbox{}itted values as the optimal solution. In this case, 
it is recommended that a robust f\mbox{}it be made to the data.
\item Fit the data in columns (i.e., variable $\lambda$, f\mbox{}ixed $N$) and `f\mbox{}ill in' the values at the places which contained the (removed) outliers. One 
may then search for outliers in the $N$ direction (i.e., for f\mbox{}ixed $\lambda$), with the application of Rosner's test, as described earlier. The estimated 
average values and accompanying uncertainties may be submitted to the f\mbox{}inal f\mbox{}it, for the creation of a smooth solution at the given ($\alpha$, $\beta$, 
$\lambda_0$) combination.
\item The third case combines the two aforementioned options and, because of this, it may be more reliable. In order to obtain an impression of 
the overall trend of the data, one starts with a (robust) f\mbox{}it to all values. One then applies the second option, using the obtained trend as guideline whenever 
a decision must be made when performing Rosner's test (e.g., in case of dif\mbox{}ferent numbers of outliers for the three signif\mbox{}icance levels).
\end{itemize}
Despite the fact that we have followed the second option in the analysis, we have nevertheless compared, at the final step, our solution with the results of the robust 
f\mbox{}it to the data.

A few comments concerning the form of the functions, which have been used in the f\mbox{}its, are due. Given the absence of theoretical guideline regarding the 
$\lambda$-dependence of the angle $\omega$, our objective is simply to choose forms which may ef\mbox{}f\mbox{}iciently achieve the data description. To this end, maximal 
freedom must be rendered to the data by choosing forms with many adjustable parameters; in such cases, the rule of thumb usually is that the introduction 
of one additional parameter should not bring noticeable improvement of the data description (there exist accurate tests, via the comparison of the $p$-values 
of the corresponding f\mbox{}its). To enable a judgement on the quality of the reproduction of the data, the original data will also be shown in almost all the f\mbox{}igures 
in Section \ref{sec:Results}. We will now comment on the empirical forms we have chosen.
\begin{itemize}
\item $\alpha<-0.05$. It is evident from the inspection of the $\omega$ values that a curve, which is capable of describing data with negative curvature, must 
be f\mbox{}itted to the data. We have tested a few forms and f\mbox{}inally decided to make use of the sum of an exponential and a $\Gamma$ function (six parameters in total).
\item $-0.05 \leq \alpha \leq 0.05$. There is some curvature in the $\alpha=-0.05$ data in the small-$\lambda$ region. The data at $\alpha=0.05$ do not 
show signif\mbox{}icant curvature. In both cases, a cubic f\mbox{}it to the logarithm of the $\omega$ values was found adequate.
\item $\alpha>0.05$. This was the most demanding case as the data show a sharp increase for decreasing $\lambda$. The f\mbox{}its were performed on the basis of a 
six-degree polynomial to the extracted $\omega$ values. A few other functions were also tested, but yielded inferior results.
\end{itemize}

We will now comment on a number of additional technical points. The integrations yielding the terms $I_i$ (given in Eqs.~(\ref{eq:I1})-(\ref{eq:I4})) 
were performed by using the trapezoidal rule (with the tolerance level set to $10^{-6}$). Other methods were tried (e.g., Romberg's integration method), but were 
found slow and did not yield more accurate results. The stability (in the series of the extracted $\omega$ values) increases substantially, if the integrations relating to 
$I_1 (u,k)$ and $I_2 (u,k)$ are performed within the roots of the Chebyshev polynomials. The values of the functions $Q_{1,2} (\zeta)$ and $R_{1,2} (\zeta)$ (see Eqs.~(C13) of 
\cite{SH}) were obtained by using the (fast) Gauss-Legendre quadrature with $256$ nodes ($1,024$ nodes were also used without improving the results). We also applied 
the Hurwitz-Zweifel method \cite{Hurwitz1,Hurwitz2}, but could not obtain accurate enough results in the problem. It is essential in the evaluation to set the regions 
of importance of the functions $F_{1,2} (\zeta)$ properly. Adaptive integration methods should (in principle) f\mbox{}ind fertile ground for application in this subject, yet 
(despite the ef\mbox{}fort) we have not been able to set up such a scheme in a direct way. Instead, we have implemented an indirect way of incorporating adaptivity, by setting 
the step in the tabulation of $F_{1,2} (\zeta)$ in such a way as to account for the importance (largeness) of the corresponding values of (each of) these functions. 
Hence, the very detailed (high-resolution) step (of $10^{-4}$ in $\zeta$) was used in the important regions (def\mbox{}ined as those regions where the value of the 
corresponding function exceeds $1 \%$ of the global maximum), whereas the low-resolution step (of $10^{-2}$ in $\zeta$) was used below that level. Contributions, 
falling below the (arbitrary) $10^{-6}$ level irrevocably, were not pursued further (thus, the tails of the integrands make no contributions to the evaluation 
of the terms $I_i$ of Eqs.~(\ref{eq:I1})-(\ref{eq:I4})).

\subsubsection{\label{sec:Smoothing}Optimisation}

For the purpose of f\mbox{}itting, the standard MINUIT package \cite{James} of the CERN library was used; to be precise, we have used the C++ version of the library 
\cite{JamesWinkler}. Extensive information on this software package may be obtained from the internet \cite{MINUITinfo}. The latest release of the code is 
available from \cite{MINUITdownload}. Version $5.28.00$ of Minuit2 (i.e., the current version at the moment we set out on this program) was incorporated in the analysis 
framework. Each optimisation was achieved on the basis of the (robust) SIMPLEX-MINIMIZE-MIGRAD chain. All f\mbox{}its terminated successfully.

The optimisation of the data description was achieved via the minimisation of a standard $\chi^2$ function; $\chi^2$ is def\mbox{}ined as the sum of the squares of the 
normalised residuals over all input data points. For each data point, the normalised residual is def\mbox{}ined as the dif\mbox{}ference between the raw (experimental) 
and f\mbox{}itted (theoretical) values, divided by the uncertainty of the `measurement'. The f\mbox{}itted values are obtained on the basis of a parametric model, for 
f\mbox{}ixed values of the model parameters; evidently, $\chi^2$ is a function of the parameter vector. Technically, the optimisation terminates when, by the 
elaborate variation of the parameter vector, the global minimum of the minimisation function is reached (more precisely, when the expected distance to the global minimum 
falls below a user-def\mbox{}ined threshold, expressed via the setting of the accuracy level in the optimisation).

For the robust f\mbox{}its, mentioned in Section \ref{sec:Extraction}, another minimisation function was chosen, namely one which is def\mbox{}ined as the sum of the 
terms $\ln (1+r_i^2)$, where $r_i$ is the normalised residual of the $i$-th data point. Compared to the standard $\chi^2$ function, this form is signif\mbox{}icantly 
less sensitive to the presence of outliers.

\subsubsection{\label{sec:BetaCorrection}$\beta$ correction}

Unlike Ref.~\cite{SH}, we do not intend to give values of the angle $\omega$ (as a function of $\lambda$) in some $\beta \neq 0$ cases at f\mbox{}ixed values 
of $\alpha$. Instead, we will give optimal values for the factor $c(\alpha, \lambda_0^{-1}; \lambda)$ def\mbox{}ined as
\begin{equation} \label{eq:DefinitionOfc}
c(\alpha, \lambda_0^{-1}; \lambda) = \frac{\omega (\alpha, \beta, \lambda_0^{-1}; \lambda) - \omega (\alpha, 0, \lambda_0^{-1}; \lambda)}{\beta} \, ,
\end{equation}
thus enabling the straightforward evaluation of $\omega (\alpha, \beta, \lambda_0^{-1}; \lambda)$ from the tabulated values of $\omega (\alpha, 0, \lambda_0^{-1}; 
\lambda)$. As it turned out that the constants $c(\alpha, \lambda_0^{-1}; \lambda)$ were dif\mbox{}ferent for $\beta<0$ and $\beta>0$, two factors per ($\alpha$, 
$\lambda_0$, $\lambda$) case will be given. The creation of these values also involves a step in which f\mbox{}itting empirical functions to the data, identif\mbox{}ied 
as the right-hand side of Eq.~(\ref{eq:DefinitionOfc}), was performed using (again) the MINUIT library.

\section{\label{sec:Results}Results}

We will now present our results for a number of values of $\lambda \leq \lambda_0/2$. The values of $\lambda_0^{-1}$ treated in the present study are: $0.10$, 
$0.25$, $0.50$, and $1.00$. We will compare our results for $\lambda_0^{-1} = 0.10$ with those obtained in Ref.~\cite{SH} (which reports no results for 
$\lambda_0^{-1}>0.10$). Given the occasional instability at $\alpha=0$ (especially at small values of $\lambda$~\footnote{The increasing instability as 
$\lambda$ approaches $0$ may have been the reason that Suo and Hutchinson \cite{SH} give no solution below $\lambda=1.0$ in the $\lambda_0^{-1}=0.05$ case.}), 
the values of the parameter $\alpha$ used were: $\pm 0.6$, $\pm 0.4$, $\pm 0.2$, and $\pm 0.05$. For the f\mbox{}irst three cases and for $\alpha=0$, Ref.~\cite{SH} 
has reported results; additionally, Ref.~\cite{SH} has extracted the $\omega$ values at $\alpha = \pm 0.8$~\footnote{For plane strain and silicon substrate, 
the $\alpha$ values when using Al, Cu, Ag, and Ti f\mbox{}ilms are: $-0.37$, $-0.09$, $-0.30$, and $-0.13$, i.e., well within the $\alpha$ interval of this work.}). 
The values of the parameter $\beta$ are subjected to the condition $\frac{\alpha-1}{4} \leq \beta \leq \frac{\alpha+1}{4}$ \cite{SH}; two values close to the upper 
and lower limits, as well as the reference value of $\beta=0$, were used.

It must be mentioned that the extraction of the $\omega$ values does not occur without ef\mbox{}fort. For instance, the $\lambda_0^{-1} = 0.10$ case will involve 
$624$ combinations of the parameter values, each combination (run) requiring anything between $30$ and $90$ minutes on a fairly-fast computer, i.e., an overall 
time load of no less than about three weeks. The $\omega$ values were produced with a dedicated C/C++-based $64$-bit application, developed within the Microsoft 
Visual Studio $9.0$ framework and run on an Intel$^\copyright$ Core$^{\rm TM}2$ Duo Processor T$9600$ at $2.80$GHz.

\begin{figure}
\begin{center}
\includegraphics [angle=270,width=15.5cm] {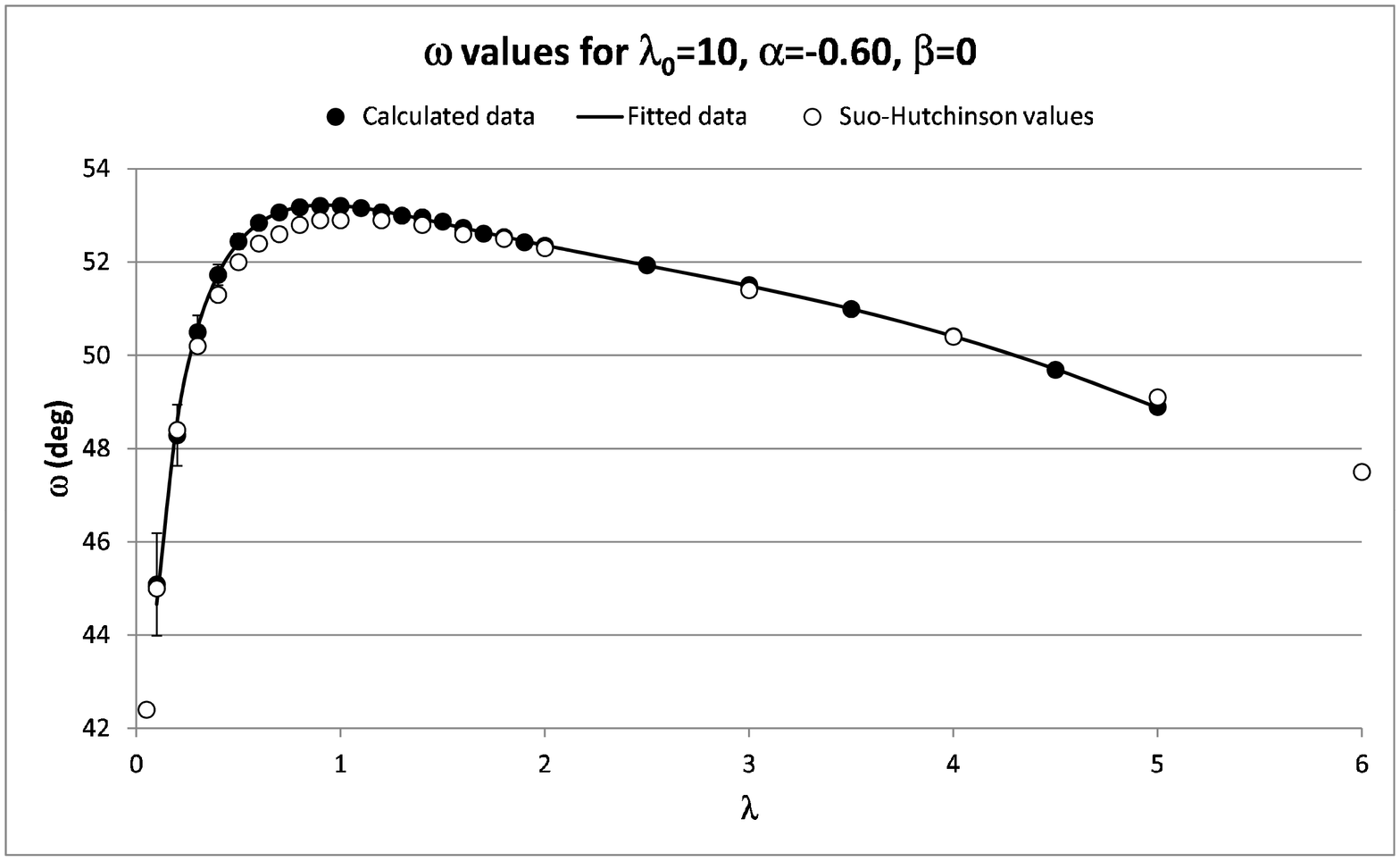}
%\vspace{-6cm}
\caption{\label{fig:Omega0p10M0p60}The $\omega$ values extracted for ($\alpha$, $\beta$, $\lambda_0^{-1}$) $=$ ($-0.60$, $0$, $0.10$) as a function of $\lambda$. 
The results of Suo and Hutchinson \cite{SH} are shown for comparison. Explicit uncertainties have not been given in Ref.~\cite{SH}.}
\end{center}
\end{figure}

\begin{figure}
\begin{center}
\includegraphics [angle=270,width=15.5cm] {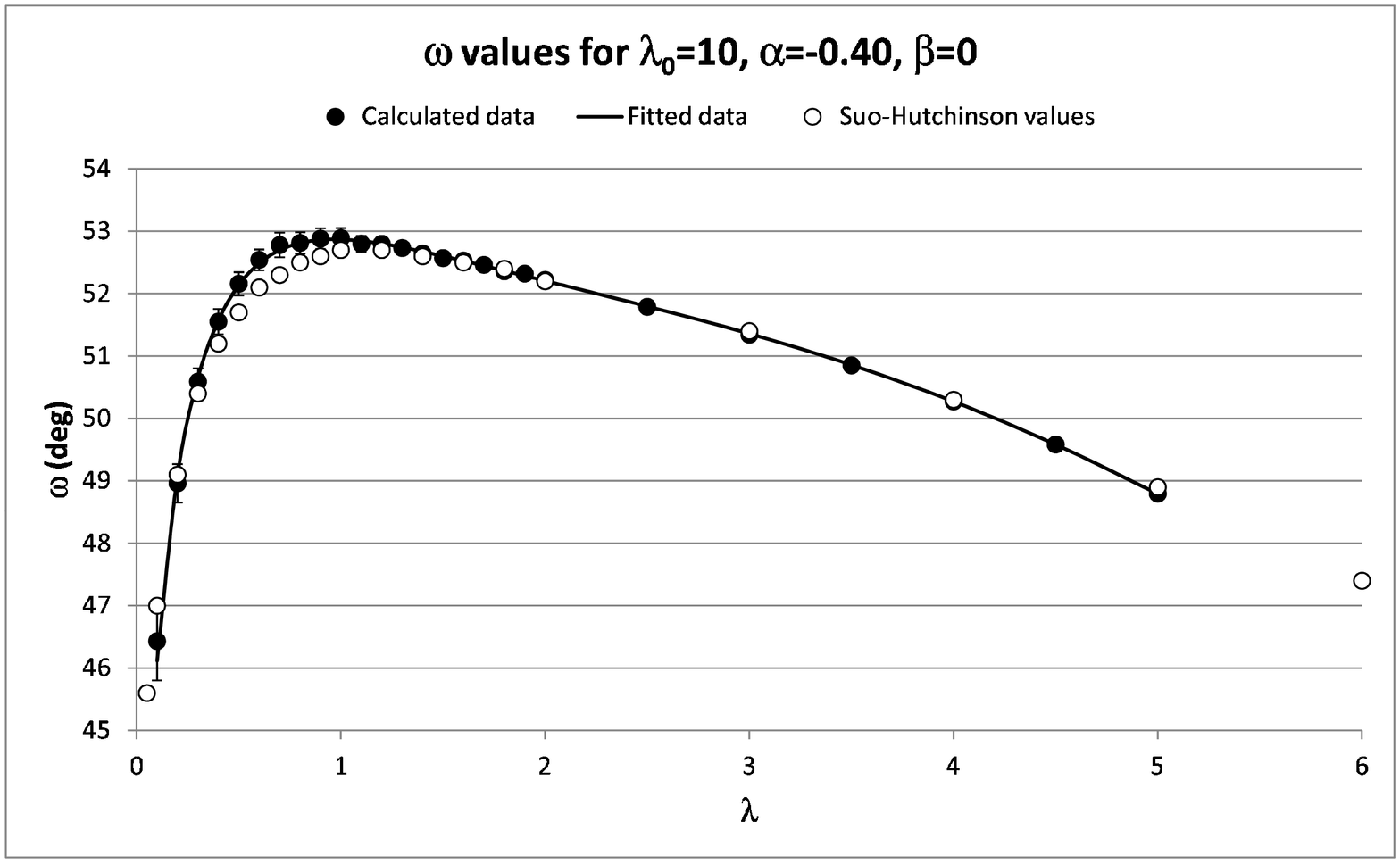}
%\vspace{-6cm}
\caption{\label{fig:Omega0p10M0p40}The $\omega$ values extracted for ($\alpha$, $\beta$, $\lambda_0^{-1}$) $=$ ($-0.40$, $0$, $0.10$) as a function of $\lambda$. 
The results of Suo and Hutchinson \cite{SH} are shown for comparison. Explicit uncertainties have not been given in Ref.~\cite{SH}.}
\end{center}
\end{figure}

\begin{figure}
\begin{center}
\includegraphics [angle=270,width=15.5cm] {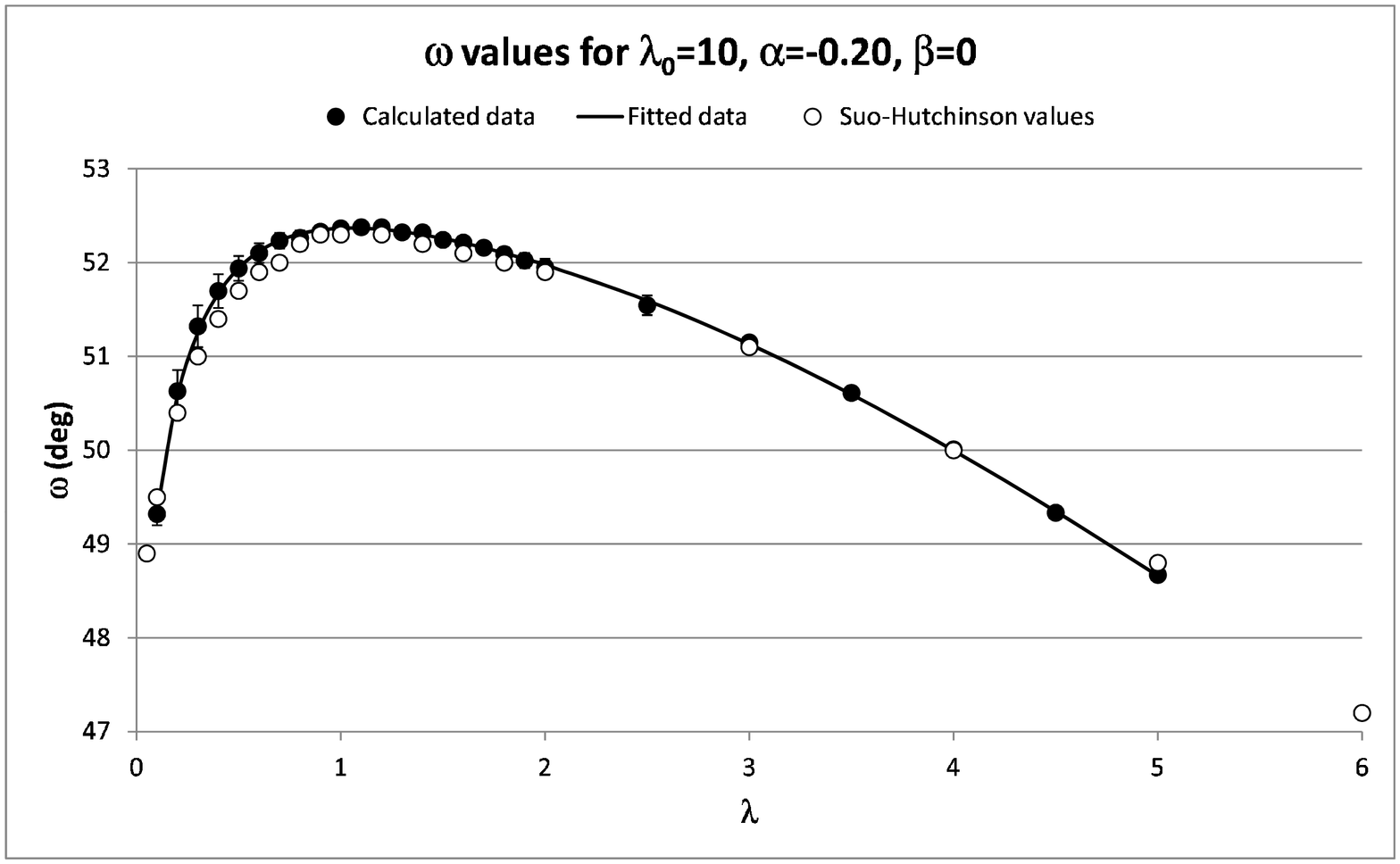}
%\vspace{-6cm}
\caption{\label{fig:Omega0p10M0p20}The $\omega$ values extracted for ($\alpha$, $\beta$, $\lambda_0^{-1}$) $=$ ($-0.20$, $0$, $0.10$) as a function of $\lambda$. 
The results of Suo and Hutchinson \cite{SH} are shown for comparison. Explicit uncertainties have not been given in Ref.~\cite{SH}.}
\end{center}
\end{figure}

\begin{figure}
\begin{center}
\includegraphics [angle=270,width=15.5cm] {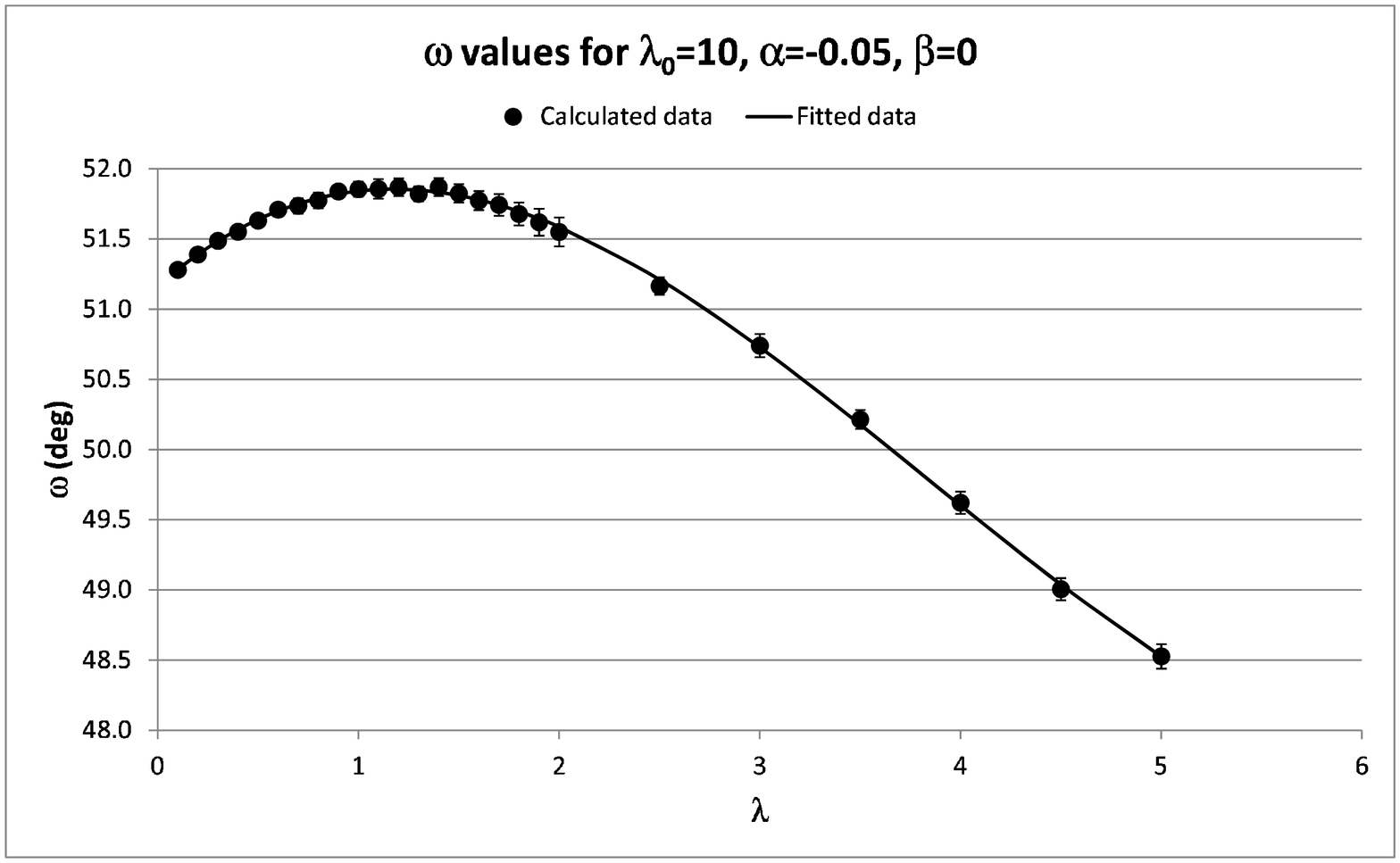}
%\vspace{-6cm}
\caption{\label{fig:Omega0p10M0p05}The $\omega$ values extracted for ($\alpha$, $\beta$, $\lambda_0^{-1}$) $=$ ($-0.05$, $0$, $0.10$) as a function of $\lambda$.}
\end{center}
\end{figure}

\begin{figure}
\begin{center}
\includegraphics [angle=270,width=15.5cm] {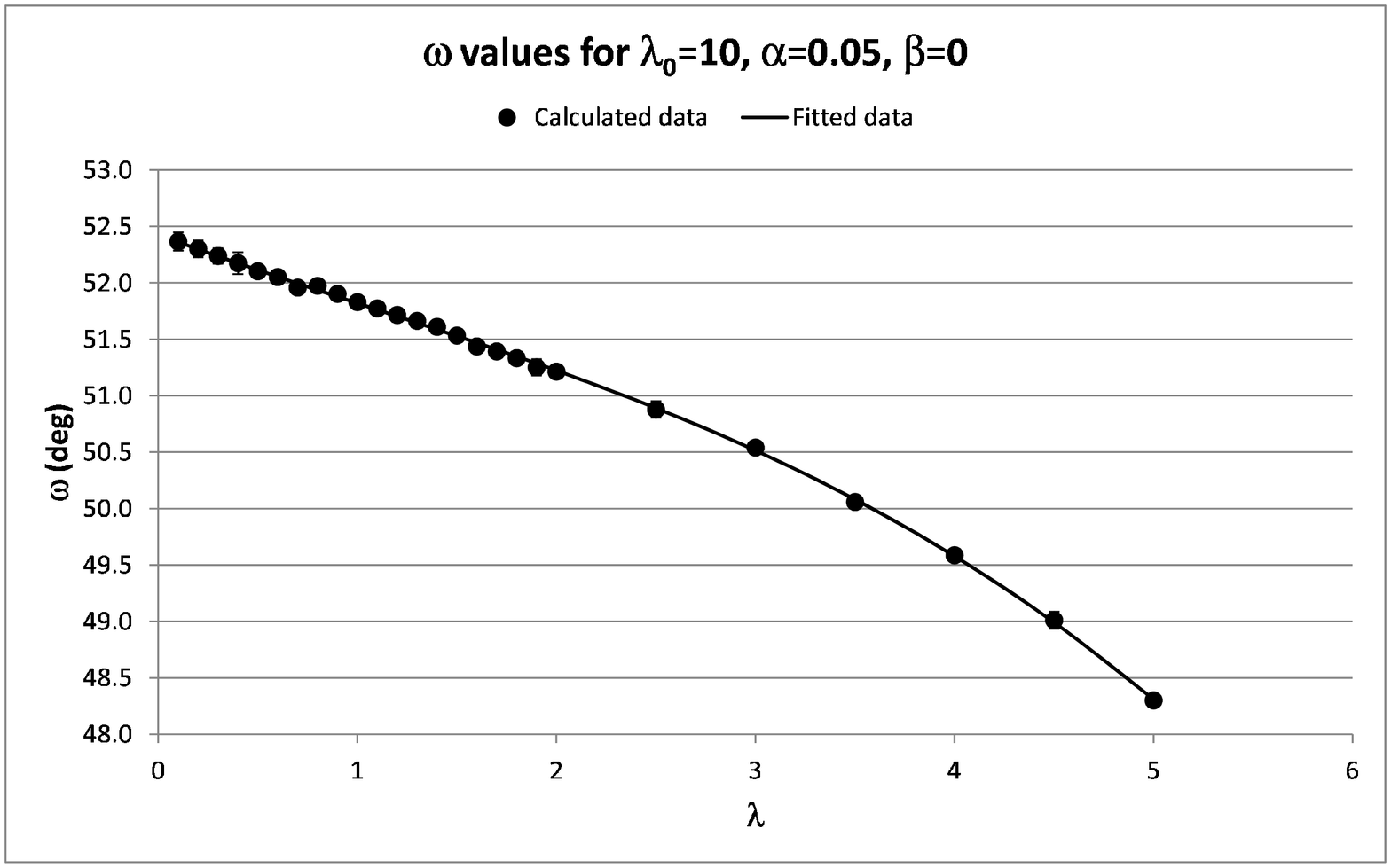}
%\vspace{-6cm}
\caption{\label{fig:Omega0p10P0p05}The $\omega$ values extracted for ($\alpha$, $\beta$, $\lambda_0^{-1}$) $=$ ($0.05$, $0$, $0.10$) as a function of $\lambda$.}
\end{center}
\end{figure}

\begin{figure}
\begin{center}
\includegraphics [angle=270,width=15.5cm] {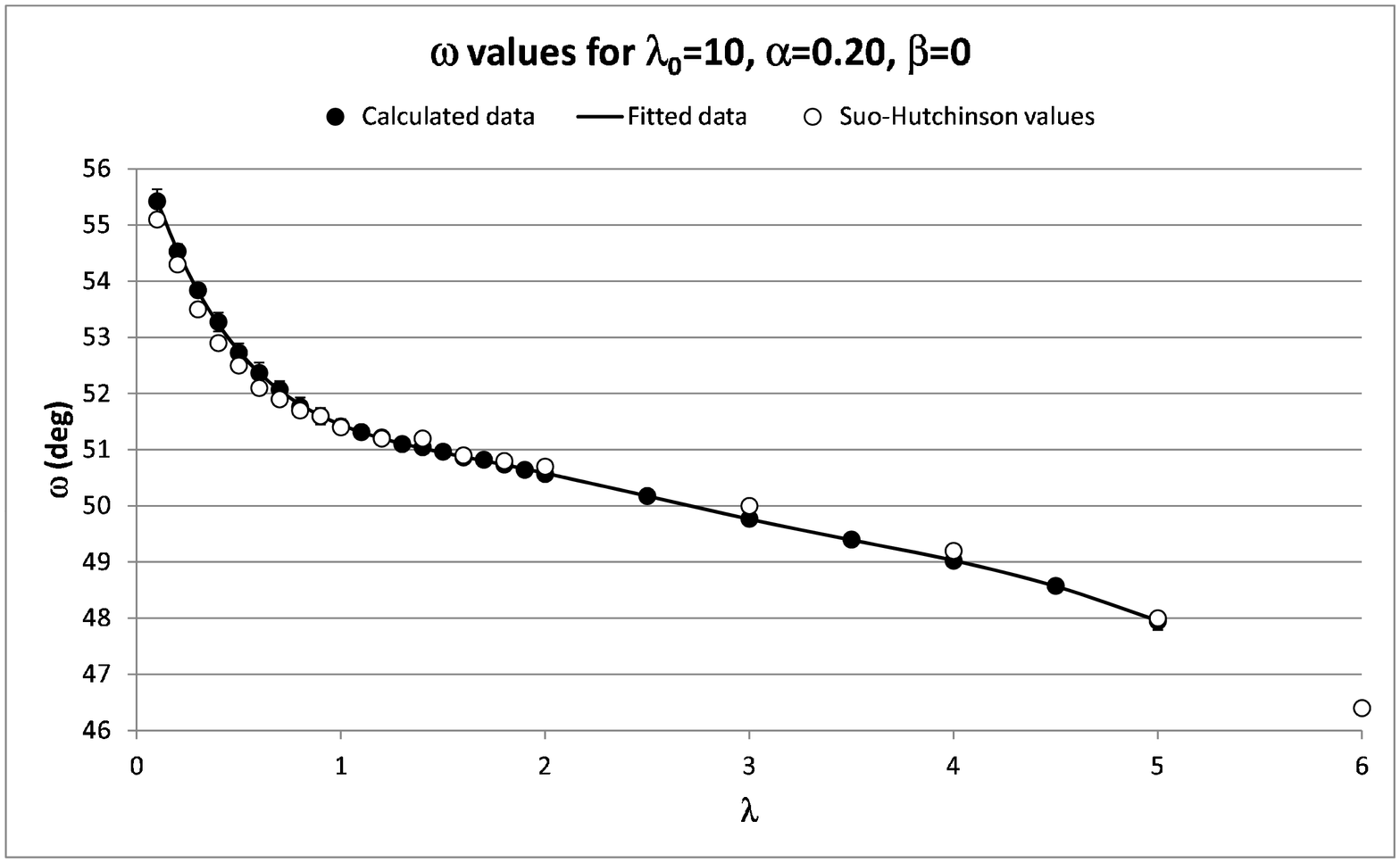}
%\vspace{-6cm}
\caption{\label{fig:Omega0p10P0p20}The $\omega$ values extracted for ($\alpha$, $\beta$, $\lambda_0^{-1}$) $=$ ($0.20$, $0$, $0.10$) as a function of $\lambda$. 
The results of Suo and Hutchinson \cite{SH} are shown for comparison. Explicit uncertainties have not been given in Ref.~\cite{SH}.}
\end{center}
\end{figure}

\begin{figure}
\begin{center}
\includegraphics [angle=270,width=15.5cm] {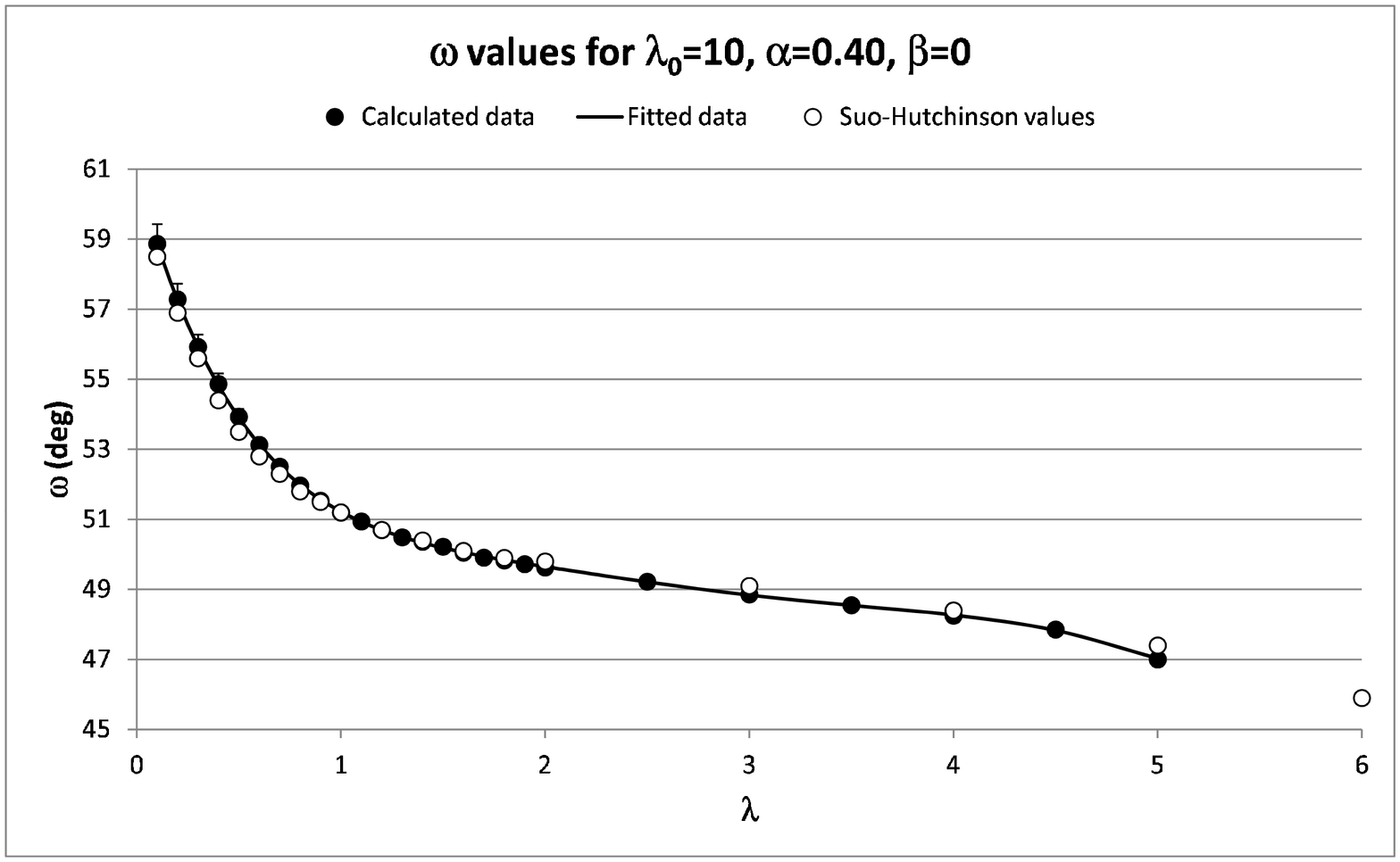}
%\vspace{-6cm}
\caption{\label{fig:Omega0p10P0p40}The $\omega$ values extracted for ($\alpha$, $\beta$, $\lambda_0^{-1}$) $=$ ($0.40$, $0$, $0.10$) as a function of $\lambda$. 
The results of Suo and Hutchinson \cite{SH} are shown for comparison. Explicit uncertainties have not been given in Ref.~\cite{SH}.}
\end{center}
\end{figure}

\begin{figure}
\begin{center}
\includegraphics [angle=270,width=15.5cm] {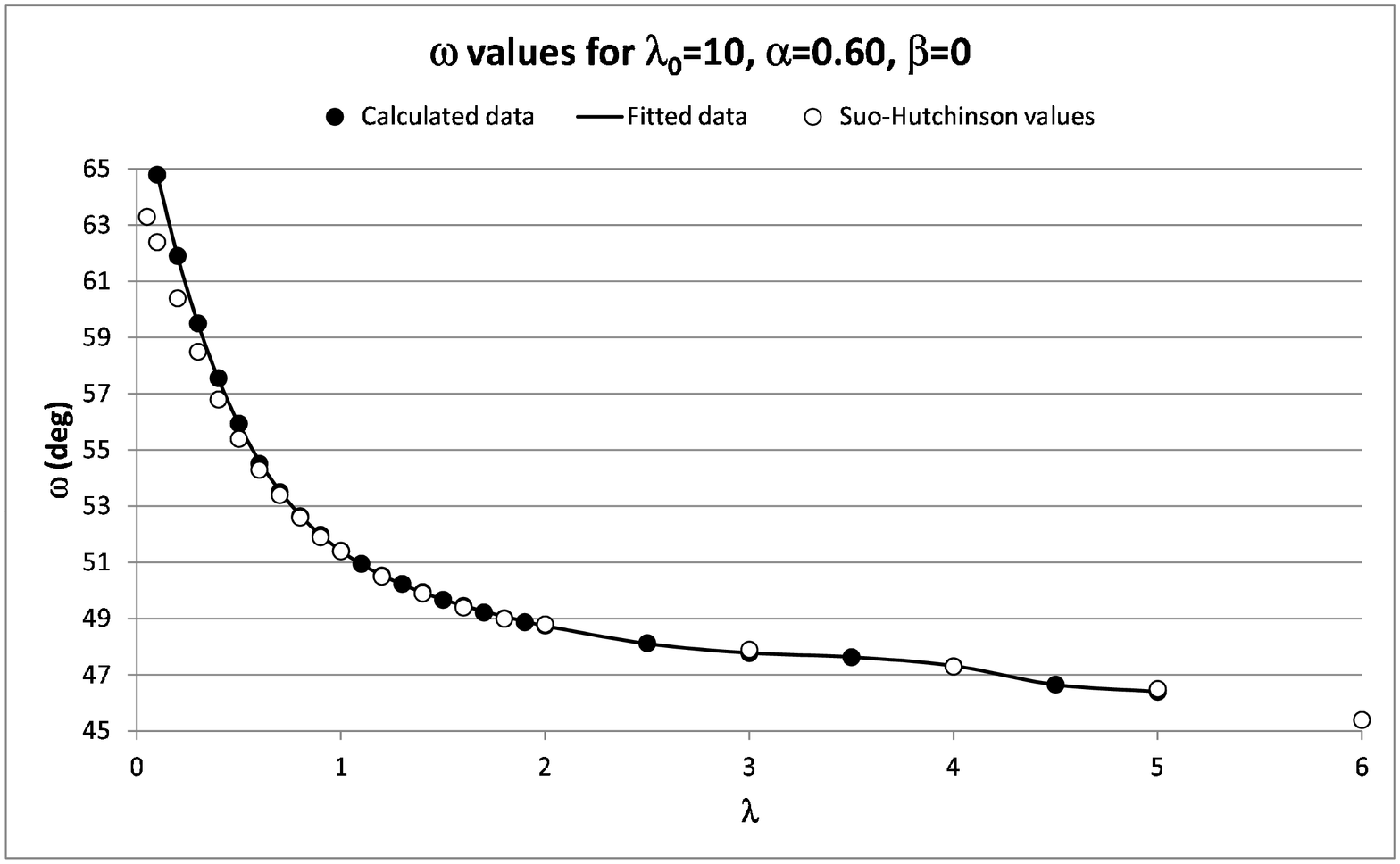}
%\vspace{-6cm}
\caption{\label{fig:Omega0p10P0p60}The $\omega$ values extracted for ($\alpha$, $\beta$, $\lambda_0^{-1}$) $=$ ($0.60$, $0$, $0.10$) as a function of $\lambda$. 
The results of Suo and Hutchinson \cite{SH} are shown for comparison. Explicit uncertainties have not been given in Ref.~\cite{SH}.}
\end{center}
\end{figure}

We will f\mbox{}irst give our results for $\lambda_0^{-1}=0.10$. Our solution for $\omega$ as a function of $\lambda$ is given in Table \ref{tab:Omega0p10} (for the 
eight $\alpha$ values used herein). The comparison of the results of this work with the Suo-Hutchinson values~\footnote{By plotting the Suo-Hutchinson data 
(versus $\lambda$), one cannot fail noticing that the values must have been subjected to some smoothing; they seem too `good' to have been obtained 
directly. We have not found comments on this issue in Ref.~\cite{SH}.} may be found in Figs.~\ref{fig:Omega0p10M0p60}-\ref{fig:Omega0p10P0p60}. We observe 
that the two sets of data agree well, save for a few values in the small-$\lambda$ region. Small dif\mbox{}ferences are also observed around $\lambda=1.0$ for $\alpha<0$, 
but die of\mbox{}f rapidly with increasing $\lambda$.

For $\alpha=-0.60$, our f\mbox{}itted solution slightly exceeds the values reported in Ref.\cite{SH} in the region $0.3 \leq \lambda \leq 1.0$, by about $0.40^0$. For 
$\alpha=-0.40$, the Suo-Hutchinson solution exceeds ours at small $\lambda$, by about $0.9^0$ at $\lambda=0.1$; the two solutions cross each other between 
$\lambda=0.2$ and $0.3$, after which our solution is larger by no more than about $0.45^0$, until the dif\mbox{}ference drops below $0.1^0$ above $\lambda=1.0$. A similar 
trend may be observed for $\alpha=-0.20$. For $\alpha>0$, the shapes of the two solutions are slightly dif\mbox{}ferent in the small-$\lambda$ region, our values exceeding 
those of Suo and Hutchinson by a few tenths of one degree. Evidently, the most serious discrepancy occurs at $\alpha=0.6$; all $\omega$ values between $\lambda=0.1$ 
and $0.6$ strongly disagree, the largest discrepancy ($2.4^0$) occurring at $\lambda=0.1$. The dif\mbox{}ferences fall below $0.1^0$ above $\lambda=0.7$. Despite the ef\mbox{}fort, 
we have not been able to pin down the source of this discrepancy, which appears to be persistent irrespective of the treatment of the data and method of outlier 
detection and removal.

We have compared our solution with the one obtained on the basis of the robust f\mbox{}it to the data after the removal of only the obvious outliers. Given that the 
dif\mbox{}ference to the Suo-Hutchinson values is signif\mbox{}icant only when $\alpha=0.6$, we will report only in that case. The result of the robust f\mbox{}it to the 
data ($93$ data points in total) gives slightly dif\mbox{}ferent values for $\lambda=0.1$, $0.2$, and $0.3$; the values obtained are: $64.66$, $61.78$, and $59.41^0$, 
respectively; above $\lambda=0.3$, the dif\mbox{}ferences between the two solutions remain below $0.1^0$. Evidently, the robust-f\mbox{}it solution conf\mbox{}irms the results of the 
elaborate analysis and thorough outlier rejection as described in Section \ref{sec:Technical}.

The two sets of the $\beta$-correction factors $c(\alpha, \lambda_0^{-1}; \lambda)$, def\mbox{}ined in Eq.~(\ref{eq:DefinitionOfc}), for the $\lambda_0^{-1}=0.10$ case are given 
in Tables \ref{tab:CorrectionForOmega0p10p} and \ref{tab:CorrectionForOmega0p10n}. Both factors show peaks below $\lambda=1.0$. For negative values of $\beta$, 
the maximal correction is around $3.40^0$ at $\alpha=-0.6$, dropping down to $1.21^0$ at $\alpha=0.6$; the minimal correction remains above $-0.37^0$. For positive values 
of $\beta$ the corrections lie between $-2.92$ and $2.37^0$. Of course, the limits given in this paragraph are linked to the domain of the $\alpha$ values used in the present work.

The detailed comparison with the results of Suo and Hutchinson, wherever it was possible to compare the two $\beta$ corrections, is given in Table \ref{tab:BetaCorrections0p10}. 
We observe that the dif\mbox{}ferences between the two solutions do not exceed a few tenths of one degree (actually, between $-0.32$ and $0.30^0$), with an average value of 
$0.02^0$ and a root-mean-square of $0.11^0$.

\begin{figure}
\begin{center}
\includegraphics [angle=270,width=15.5cm] {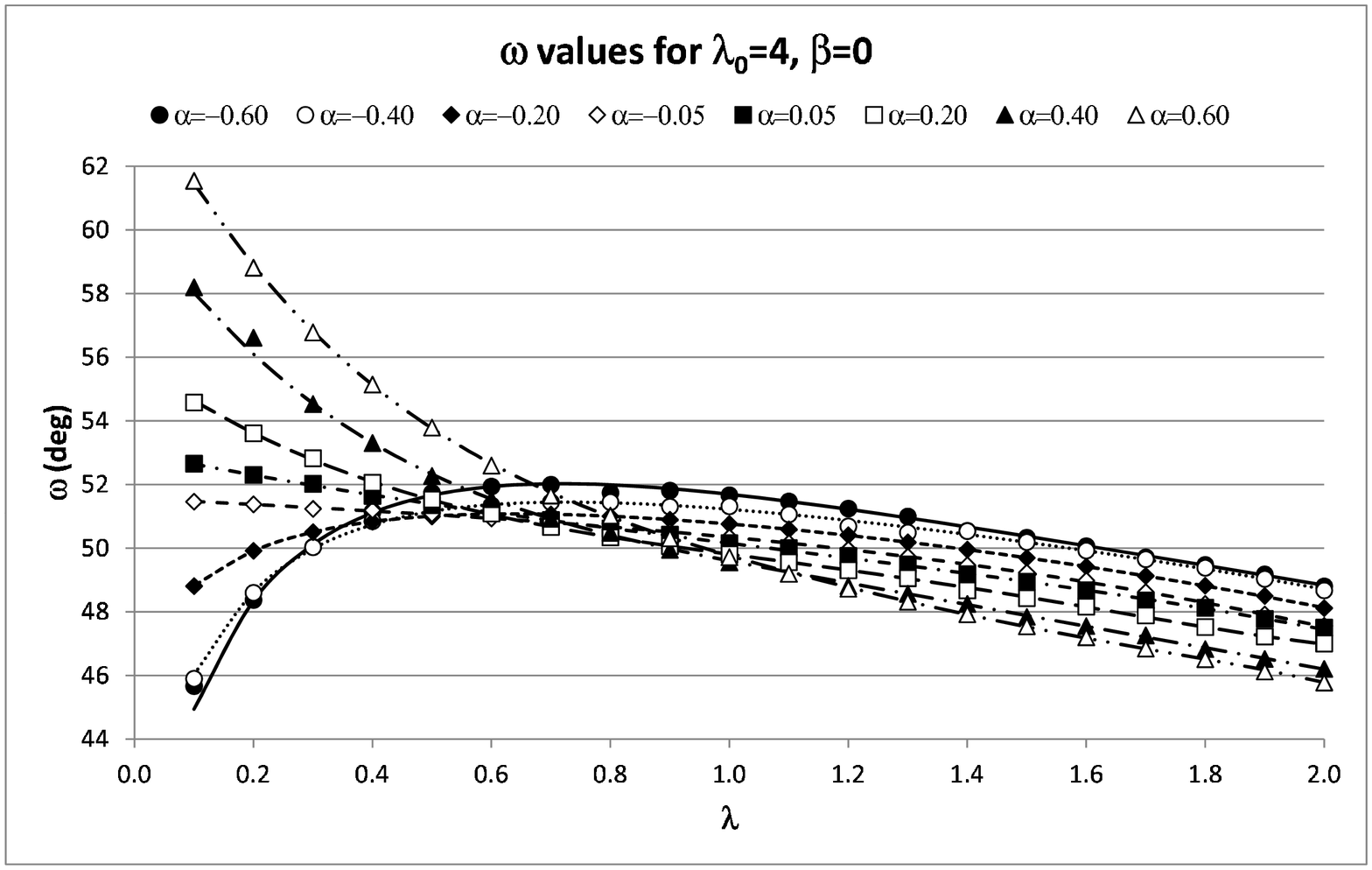}
%\vspace{-6cm}
\caption{\label{fig:Omega0p25}The $\omega$ values extracted for ($\beta$, $\lambda_0^{-1}$) $=$ ($0$, $0.25$) as a function of $\lambda$ for all the values 
of the parameter $\alpha$ used in this work. For reasons of clarity, the uncertainties in the data are not shown. The f\mbox{}itted curves, yielding the data of 
Table \ref{tab:Omega0p25}, are also shown.}
\end{center}
\end{figure}

\begin{figure}
\begin{center}
\includegraphics [angle=270,width=15.5cm] {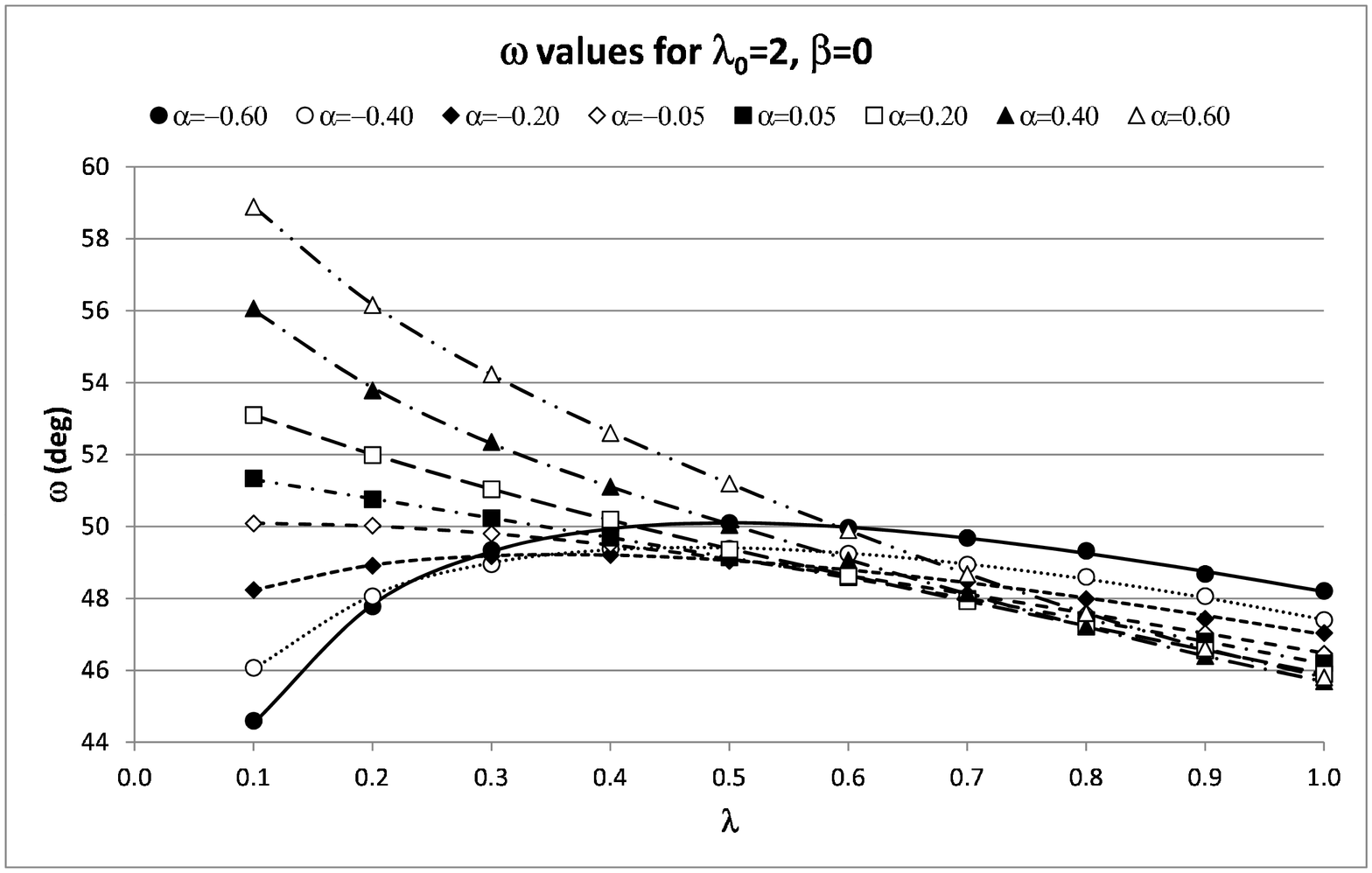}
%\vspace{-6cm}
\caption{\label{fig:Omega0p50}The $\omega$ values extracted for ($\beta$, $\lambda_0^{-1}$) $=$ ($0$, $0.50$) as a function of $\lambda$ for all the values 
of the parameter $\alpha$ used in this work. For reasons of clarity, the uncertainties in the data are not shown. The f\mbox{}itted curves, yielding the data of 
Table \ref{tab:Omega0p50}, are also shown.}
\end{center}
\end{figure}

\begin{figure}
\begin{center}
\includegraphics [angle=270,width=15.5cm] {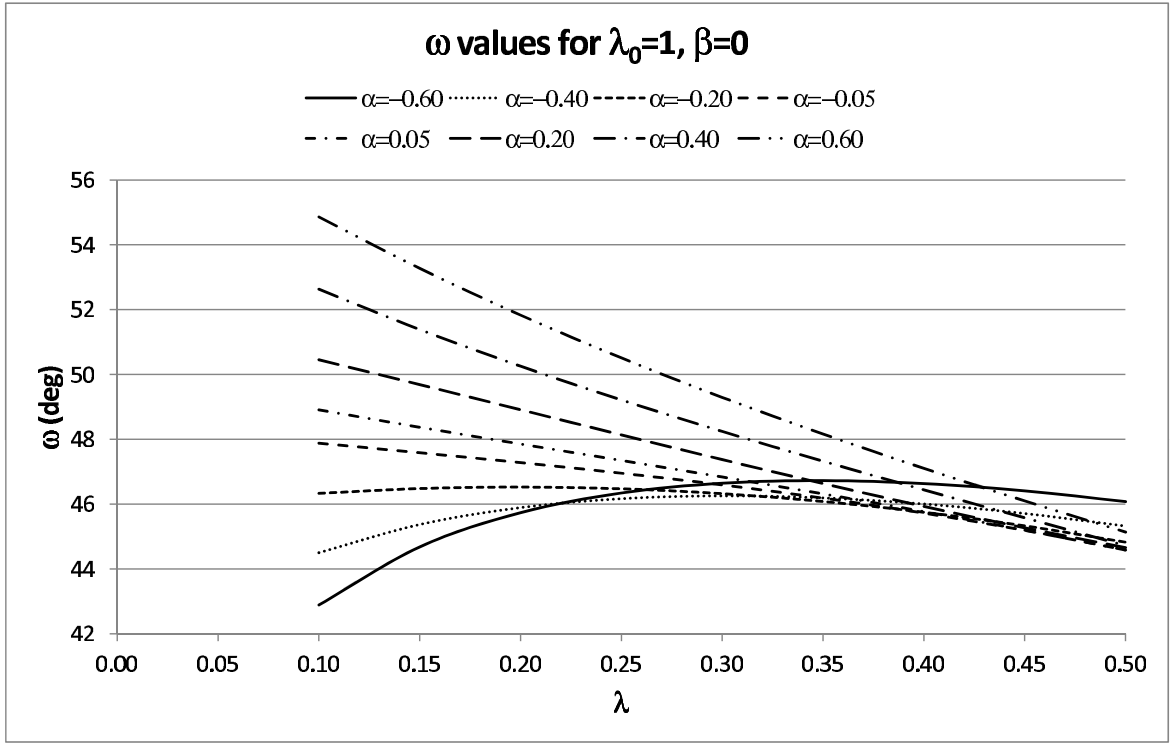}
%\vspace{-6cm}
\caption{\label{fig:Omega1p00}The $\omega$ values extracted for ($\beta$, $\lambda_0^{-1}$) $=$ ($0$, $1.00$) as a function of $\lambda$ for all the values 
of the parameter $\alpha$ used in this work. Only the f\mbox{}itted curves, yielding the data of Table \ref{tab:Omega1p00}, are shown; it does not make sense 
to show also the original data in this case, as they almost lie on the displayed curves.}
\end{center}
\end{figure}

Our results for the angle $\omega$ for $\lambda_0^{-1}=0.25$, $0.50$, and $1.00$ are to be found in Tables \ref{tab:Omega0p25}, \ref{tab:Omega0p50}, and \ref{tab:Omega1p00}, 
respectively. The corresponding correction factors $c(\alpha, \lambda_0^{-1}; \lambda)$, def\mbox{}ined in Eq.~(\ref{eq:DefinitionOfc}), are listed in Tables 
\ref{tab:CorrectionForOmega0p25}, \ref{tab:CorrectionForOmega0p50}, and \ref{tab:CorrectionForOmega1p00}. Concerning these cases, the only interesting remark is that the 
extracted $\omega$ values showed less f\mbox{}luctuation than in the $\lambda_0^{-1}=0.10$ case. All data have been obtained by applying the process outlined in Section 
\ref{sec:Extraction} and have been subjected to smoothing as described in Section \ref{sec:Smoothing}. In schematic form, the data (along with the f\mbox{}itted curves) are 
displayed in Figs.~\ref{fig:Omega0p25}-\ref{fig:Omega1p00}; the uncertainties have been omitted. It is interesting to notice that the $\beta$ corrections may become important 
close to the interface of the two materials for $\alpha>0$, more important than they seem to be in the $\lambda_0^{-1}=0.10$ case.

\section{\label{sec:Conclusions}Conclusions}

Our aim in this work has been to re-address the subject of steady-state cracking, originally investigated by Suo and Hutchinson \cite{SH} more than thirty 
years ago. One of the reasons calling for further investigation is the extraction of solutions for values of the substrate-to-f\mbox{}ilm thickness ratio 
smaller than $10$, which was the lowest value treated in the Suo-Hutchinson paper. To this end, we produced the solutions for the ratio values of $4$, $2$, 
and $1$. Our solutions may be used as `anchor points' in the evaluation of the quantity $\omega$ (introduced in Ref.~\cite{SH}, to link the external mechanical 
loads and moments with the stress intensity factors) in the general problem of arbitrary thickness ratio.

We have also obtained the solution in one of the two cases treated in Ref.~\cite{SH}, namely at the substrate-to-f\mbox{}ilm thickness ratio of $10$. We have 
concluded that the two sets of $\omega$ values agree well, save for distances close to the interface of the two materials (small $\lambda$ values), where 
a signif\mbox{}icant dif\mbox{}ference (of $2.4^0$) has been found on one occasion.

We have not been able to observe true convergence in the series of the extracted $\omega$ values when increasing the dimensionality of the problem, i.e., the 
number of Chebyshev polynomials used in Eq.~(\ref{eq:B12}). In the absence of convergence, we have decided to apply an algorithm for the detection of outliers, 
estimate the average values (and accompanying uncertainties) of the surviving data points, and f\mbox{}it simple forms to the resulting (`clean') data. We 
present our results in tabular form, thus enabling their easy implementation.

One of the important contributions of the present work is the introduction of a well-ordered scheme for the application of the $\beta$ correction. Essential, in 
this respect, was the observation that this correction scales with $\beta$, yet differently for positive and negative $\beta$ values. To facilitate the application 
of this correction, we have given the relevant multiplicative factors in tabular form.

\begin{ack}
The author acknowledges helpful discussions with Peter Biller.
% Concerning the application of the results of the present study, the author acknowledges stimulating 
%discussions with Felix Budde, Raja Dravid, Lukas Lichtensteiger, Max Lungarella, and Hau-Kit Man of Enexra Tools GmbH. This work has been f\mbox{}inancially supported 
%(in part) by the `CTI - The Innovation Promotion Agency' of the `Federal Of\mbox{}f\mbox{}ice for Professional Education and Technology (OPET)'; project title: `Semi-automatic 
%robotic system for kerf-free silicon wafer production', KTI-Projekt Nr. 11156.2.
\end{ack}

\newpage
\begin{table}[h!]
{\bf \caption{\label{tab:Omega0p10}}}The values of the angle $\omega$ (in degrees) for $\lambda_0^{-1}=0.10$ and $\beta=0$.
\vspace{0.2cm}
\begin{center}
\begin{tabular}{|c|c|c|c|c|c|c|c|c|}
\hline
$\lambda \downarrow$, $\alpha \rightarrow$ & $-0.60$ & $-0.40$ & $-0.20$ & $-0.05$ & $0.05$ & $0.20$ & $0.40$ & $0.60$ \\
\hline
$0.1$ & $44.66$ & $46.12$ & $49.33$ & $51.28$ & $52.36$ & $55.43$ & $58.83$ & $64.81$ \\
$0.2$ & $48.57$ & $49.08$ & $50.57$ & $51.39$ & $52.30$ & $54.55$ & $57.26$ & $61.91$ \\
$0.3$ & $50.57$ & $50.65$ & $51.25$ & $51.48$ & $52.24$ & $53.83$ & $55.94$ & $59.51$ \\
$0.4$ & $51.73$ & $51.57$ & $51.67$ & $51.57$ & $52.17$ & $53.24$ & $54.83$ & $57.54$ \\
$0.5$ & $52.42$ & $52.14$ & $51.94$ & $51.64$ & $52.11$ & $52.76$ & $53.91$ & $55.92$ \\
$0.6$ & $52.83$ & $52.49$ & $52.12$ & $51.70$ & $52.05$ & $52.37$ & $53.14$ & $54.61$ \\
$0.7$ & $53.06$ & $52.70$ & $52.23$ & $51.75$ & $52.00$ & $52.06$ & $52.51$ & $53.55$ \\
$0.8$ & $53.18$ & $52.81$ & $52.31$ & $51.79$ & $51.94$ & $51.81$ & $51.99$ & $52.68$ \\
$0.9$ & $53.22$ & $52.86$ & $52.35$ & $51.82$ & $51.88$ & $51.61$ & $51.57$ & $51.98$ \\
$1.0$ & $53.21$ & $52.87$ & $52.37$ & $51.84$ & $51.82$ & $51.44$ & $51.22$ & $51.41$ \\
$1.1$ & $53.16$ & $52.84$ & $52.37$ & $51.85$ & $51.76$ & $51.31$ & $50.93$ & $50.94$ \\
$1.2$ & $53.10$ & $52.80$ & $52.35$ & $51.85$ & $51.70$ & $51.20$ & $50.69$ & $50.55$ \\
$1.3$ & $53.01$ & $52.74$ & $52.33$ & $51.85$ & $51.65$ & $51.11$ & $50.50$ & $50.22$ \\
$1.4$ & $52.92$ & $52.67$ & $52.30$ & $51.83$ & $51.59$ & $51.03$ & $50.33$ & $49.93$ \\
$1.5$ & $52.83$ & $52.60$ & $52.26$ & $51.81$ & $51.53$ & $50.95$ & $50.18$ & $49.68$ \\
$1.6$ & $52.73$ & $52.53$ & $52.21$ & $51.78$ & $51.47$ & $50.88$ & $50.06$ & $49.46$ \\
$1.7$ & $52.64$ & $52.45$ & $52.16$ & $51.74$ & $51.41$ & $50.80$ & $49.94$ & $49.26$ \\
$1.8$ & $52.54$ & $52.37$ & $52.10$ & $51.70$ & $51.35$ & $50.73$ & $49.84$ & $49.07$ \\
$1.9$ & $52.45$ & $52.29$ & $52.04$ & $51.64$ & $51.29$ & $50.66$ & $49.74$ & $48.90$ \\
$2.0$ & $52.36$ & $52.21$ & $51.98$ & $51.59$ & $51.22$ & $50.58$ & $49.65$ & $48.74$ \\
$2.5$ & $51.93$ & $51.80$ & $51.60$ & $51.21$ & $50.89$ & $50.18$ & $49.21$ & $48.11$ \\
$3.0$ & $51.49$ & $51.36$ & $51.13$ & $50.73$ & $50.51$ & $49.76$ & $48.84$ & $47.78$ \\
$3.5$ & $51.00$ & $50.86$ & $50.60$ & $50.18$ & $50.08$ & $49.39$ & $48.54$ & $47.63$ \\
$4.0$ & $50.41$ & $50.27$ & $50.00$ & $49.60$ & $49.58$ & $49.03$ & $48.27$ & $47.32$ \\
$4.5$ & $49.70$ & $49.58$ & $49.35$ & $49.04$ & $48.99$ & $48.57$ & $47.83$ & $46.65$ \\
$5.0$ & $48.88$ & $48.79$ & $48.66$ & $48.52$ & $48.31$ & $47.96$ & $47.03$ & $46.40$ \\
\hline
\end{tabular}
\end{center}
\end{table}

\newpage
\begin{table}[h!]
{\bf \caption{\label{tab:CorrectionForOmega0p10p}}}The $\beta$-correction factors $c(\alpha, \lambda_0^{-1}=0.10; \lambda)$ (in degrees), 
def\mbox{}ined in Eq.~(\ref{eq:DefinitionOfc}), for $\beta < 0$.
\vspace{0.2cm}
\begin{center}
\begin{tabular}{|c|c|c|c|c|c|c|c|c|}
\hline
$\lambda \downarrow$, $\alpha \rightarrow$ & $-0.60$ & $-0.40$ & $-0.20$ & $-0.05$ & $0.05$ & $0.20$ & $0.40$ & $0.60$ \\
\hline
$0.10$ & $-2.83$ & $-2.38$ & $-1.59$ & $-0.83$ & $-0.25$ & $0.71$ & $2.13$ & $3.72$ \\
$0.20$ & $-7.63$ & $-8.33$ & $-8.59$ & $-8.58$ & $-8.50$ & $-8.27$ & $-7.81$ & $-7.20$ \\
$0.30$ & $-8.49$ & $-9.56$ & $-10.21$ & $-10.52$ & $-10.66$ & $-10.78$ & $-10.79$ & $-10.66$ \\
$0.40$ & $-8.27$ & $-9.48$ & $-10.32$ & $-10.80$ & $-11.06$ & $-11.37$ & $-11.67$ & $-11.86$ \\
$0.50$ & $-7.69$ & $-8.93$ & $-9.86$ & $-10.43$ & $-10.77$ & $-11.21$ & $-11.70$ & $-12.10$ \\
$0.60$ & $-6.99$ & $-8.21$ & $-9.19$ & $-9.81$ & $-10.19$ & $-10.71$ & $-11.33$ & $-11.88$ \\
$0.70$ & $-6.28$ & $-7.46$ & $-8.44$ & $-9.09$ & $-9.50$ & $-10.07$ & $-10.77$ & $-11.43$ \\
$0.80$ & $-5.60$ & $-6.73$ & $-7.70$ & $-8.36$ & $-8.78$ & $-9.38$ & $-10.14$ & $-10.87$ \\
$0.90$ & $-4.98$ & $-6.05$ & $-6.99$ & $-7.65$ & $-8.07$ & $-8.69$ & $-9.48$ & $-10.25$ \\
$1.00$ & $-4.42$ & $-5.43$ & $-6.34$ & $-6.99$ & $-7.41$ & $-8.02$ & $-8.83$ & $-9.62$ \\
$1.10$ & $-3.93$ & $-4.87$ & $-5.74$ & $-6.37$ & $-6.78$ & $-7.39$ & $-8.20$ & $-9.01$ \\
$1.20$ & $-3.49$ & $-4.37$ & $-5.20$ & $-5.81$ & $-6.21$ & $-6.81$ & $-7.60$ & $-8.41$ \\
$1.30$ & $-3.10$ & $-3.93$ & $-4.71$ & $-5.30$ & $-5.68$ & $-6.26$ & $-7.04$ & $-7.83$ \\
$1.40$ & $-2.76$ & $-3.53$ & $-4.28$ & $-4.83$ & $-5.20$ & $-5.76$ & $-6.52$ & $-7.29$ \\
$1.50$ & $-2.46$ & $-3.18$ & $-3.88$ & $-4.41$ & $-4.77$ & $-5.30$ & $-6.03$ & $-6.78$ \\
$1.60$ & $-2.20$ & $-2.87$ & $-3.53$ & $-4.03$ & $-4.37$ & $-4.88$ & $-5.58$ & $-6.29$ \\
$1.70$ & $-1.97$ & $-2.59$ & $-3.21$ & $-3.69$ & $-4.01$ & $-4.49$ & $-5.16$ & $-5.84$ \\
$1.80$ & $-1.77$ & $-2.35$ & $-2.93$ & $-3.37$ & $-3.68$ & $-4.14$ & $-4.77$ & $-5.42$ \\
$1.90$ & $-1.59$ & $-2.13$ & $-2.67$ & $-3.09$ & $-3.38$ & $-3.81$ & $-4.41$ & $-5.03$ \\
$2.00$ & $-1.44$ & $-1.94$ & $-2.44$ & $-2.84$ & $-3.10$ & $-3.51$ & $-4.08$ & $-4.66$ \\
$2.50$ & $-0.89$ & $-1.23$ & $-1.58$ & $-1.86$ & $-2.05$ & $-2.34$ & $-2.75$ & $-3.17$ \\
$3.00$ & $-0.58$ & $-0.81$ & $-1.05$ & $-1.23$ & $-1.36$ & $-1.56$ & $-1.84$ & $-2.13$ \\
$3.50$ & $-0.38$ & $-0.54$ & $-0.70$ & $-0.82$ & $-0.91$ & $-1.04$ & $-1.22$ & $-1.42$ \\
$4.00$ & $-0.26$ & $-0.36$ & $-0.46$ & $-0.54$ & $-0.60$ & $-0.69$ & $-0.81$ & $-0.93$ \\
$4.50$ & $-0.17$ & $-0.24$ & $-0.31$ & $-0.36$ & $-0.39$ & $-0.45$ & $-0.53$ & $-0.61$ \\
$5.00$ & $-0.12$ & $-0.16$ & $-0.20$ & $-0.23$ & $-0.26$ & $-0.29$ & $-0.34$ & $-0.39$ \\
\hline
\end{tabular}
\end{center}
\end{table}

\newpage
\begin{table}[h!]
{\bf \caption{\label{tab:CorrectionForOmega0p10n}}}The $\beta$-correction factors $c(\alpha, \lambda_0^{-1}=0.10; \lambda)$ (in degrees), 
def\mbox{}ined in Eq.~(\ref{eq:DefinitionOfc}), for $\beta > 0$.
\vspace{0.2cm}
\begin{center}
\begin{tabular}{|c|c|c|c|c|c|c|c|c|}
\hline
$\lambda \downarrow$, $\alpha \rightarrow$ & $-0.60$ & $-0.40$ & $-0.20$ & $-0.05$ & $0.05$ & $0.20$ & $0.40$ & $0.60$ \\
\hline
$0.10$ & $-6.00$ & $-4.01$ & $-2.03$ & $-0.54$ & $0.45$ & $1.94$ & $3.93$ & $5.92$ \\
$0.20$ & $-8.99$ & $-7.57$ & $-6.14$ & $-5.07$ & $-4.36$ & $-3.29$ & $-1.87$ & $-0.45$ \\
$0.30$ & $-9.45$ & $-8.47$ & $-7.49$ & $-6.75$ & $-6.26$ & $-5.53$ & $-4.55$ & $-3.57$ \\
$0.40$ & $-9.18$ & $-8.55$ & $-7.91$ & $-7.44$ & $-7.12$ & $-6.65$ & $-6.01$ & $-5.38$ \\
$0.50$ & $-8.64$ & $-8.27$ & $-7.91$ & $-7.63$ & $-7.45$ & $-7.17$ & $-6.81$ & $-6.44$ \\
$0.60$ & $-7.99$ & $-7.83$ & $-7.67$ & $-7.55$ & $-7.47$ & $-7.35$ & $-7.19$ & $-7.02$ \\
$0.70$ & $-7.32$ & $-7.31$ & $-7.30$ & $-7.30$ & $-7.30$ & $-7.29$ & $-7.29$ & $-7.28$ \\
$0.80$ & $-6.66$ & $-6.76$ & $-6.87$ & $-6.95$ & $-7.01$ & $-7.09$ & $-7.19$ & $-7.30$ \\
$0.90$ & $-6.03$ & $-6.22$ & $-6.41$ & $-6.55$ & $-6.64$ & $-6.79$ & $-6.98$ & $-7.17$ \\
$1.00$ & $-5.44$ & $-5.68$ & $-5.93$ & $-6.12$ & $-6.24$ & $-6.42$ & $-6.67$ & $-6.92$ \\
$1.10$ & $-4.89$ & $-5.18$ & $-5.46$ & $-5.67$ & $-5.81$ & $-6.03$ & $-6.31$ & $-6.60$ \\
$1.20$ & $-4.39$ & $-4.70$ & $-5.00$ & $-5.23$ & $-5.39$ & $-5.61$ & $-5.92$ & $-6.23$ \\
$1.30$ & $-3.94$ & $-4.25$ & $-4.57$ & $-4.80$ & $-4.96$ & $-5.20$ & $-5.51$ & $-5.83$ \\
$1.40$ & $-3.52$ & $-3.84$ & $-4.16$ & $-4.39$ & $-4.55$ & $-4.79$ & $-5.11$ & $-5.42$ \\
$1.50$ & $-3.15$ & $-3.46$ & $-3.77$ & $-4.00$ & $-4.16$ & $-4.39$ & $-4.70$ & $-5.02$ \\
$1.60$ & $-2.81$ & $-3.11$ & $-3.41$ & $-3.64$ & $-3.79$ & $-4.01$ & $-4.32$ & $-4.62$ \\
$1.70$ & $-2.51$ & $-2.79$ & $-3.08$ & $-3.30$ & $-3.44$ & $-3.66$ & $-3.94$ & $-4.23$ \\
$1.80$ & $-2.23$ & $-2.51$ & $-2.78$ & $-2.98$ & $-3.12$ & $-3.32$ & $-3.59$ & $-3.87$ \\
$1.90$ & $-1.99$ & $-2.24$ & $-2.50$ & $-2.69$ & $-2.82$ & $-3.01$ & $-3.26$ & $-3.52$ \\
$2.00$ & $-1.77$ & $-2.01$ & $-2.24$ & $-2.42$ & $-2.54$ & $-2.72$ & $-2.96$ & $-3.19$ \\
$2.50$ & $-0.98$ & $-1.13$ & $-1.29$ & $-1.40$ & $-1.48$ & $-1.59$ & $-1.75$ & $-1.90$ \\
$3.00$ & $-0.53$ & $-0.63$ & $-0.72$ & $-0.79$ & $-0.83$ & $-0.90$ & $-0.99$ & $-1.09$ \\
$3.50$ & $-0.29$ & $-0.34$ & $-0.39$ & $-0.43$ & $-0.46$ & $-0.50$ & $-0.55$ & $-0.60$ \\
$4.00$ & $-0.15$ & $-0.18$ & $-0.21$ & $-0.23$ & $-0.25$ & $-0.27$ & $-0.30$ & $-0.33$ \\
$4.50$ & $-0.08$ & $-0.10$ & $-0.11$ & $-0.12$ & $-0.13$ & $-0.14$ & $-0.16$ & $-0.18$ \\
$5.00$ & $-0.04$ & $-0.05$ & $-0.06$ & $-0.07$ & $-0.07$ & $-0.08$ & $-0.08$ & $-0.09$ \\
\hline
\end{tabular}
\end{center}
\end{table}

\newpage
\begin{table}[h!]
{\bf \caption{\label{tab:BetaCorrections0p10}}}The comparison of the $\beta$ corrections between the Suo-Hutchinson solution \cite{SH} and those 
obtained in this work. The f\mbox{}irst three columns correspond to the values of the parameters $\alpha$, $\beta$, and $\lambda$, for which results have 
been reported in Ref.~\cite{SH}. (As our solution does not include the $\alpha$ values of $\pm 0.8$, no comparison can be made in this case.) The 
next column ($\tilde{\omega}_0$) contains the Suo-Hutchinson solution at the specif\mbox{}ic $\alpha$ and $\lambda$ values, at $\beta=0$. The adjacent 
column ($\tilde{\omega}$) contains their solution for the $\beta$ value given in the second column. The dif\mbox{}ference of the two aforementioned angles 
is shown next; this is the Suo-Hutchinson correction due to the non-zero $\beta$ value. We subsequently give our $\beta$-correction factor $c$, taken directly 
from Tables \ref{tab:CorrectionForOmega0p10p} and \ref{tab:CorrectionForOmega0p10n}, depending on the sign of $\beta$. Our correction (shown next) has 
been obtained by multiplying $\beta$ and the factor $c$. The last column contains the dif\mbox{}ference between the two $\beta$ corrections 
(i.e., our result minus the corresponding Suo-Hutchinson value). All angles (as well as the multiplicative factors $c$) are given in degrees. The 
numerical results of this table have been rounded to two decimal places.
\vspace{0.2cm}
\begin{center}
\begin{tabular}{|c|c|c|c|c|c|c|c|c|}
\hline
$\alpha$ & $\beta$ & $\lambda$ & $\tilde{\omega}_0$ & $\tilde{\omega}$ & $\tilde{\omega}-\tilde{\omega_0}$ & $c$ & $\omega-\omega_0$ & $\delta$ \\
\hline
$-0.60$ & $-0.40$ & $0.5$ & $52.00$ & $55.40$ & $3.40$ & $-7.69$ & $3.08$ & $-0.32$ \\
$-0.60$ & $-0.40$ & $1.0$ & $52.90$ & $54.80$ & $1.90$ & $-4.42$ & $1.77$ & $-0.13$ \\
$-0.60$ & $-0.40$ & $1.5$ & $52.70$ & $53.80$ & $1.10$ & $-2.46$ & $0.98$ & $-0.12$ \\
$-0.60$ & $-0.40$ & $2.0$ & $52.30$ & $52.90$ & $0.60$ & $-1.44$ & $0.58$ & $-0.02$ \\
$-0.60$ & $-0.40$ & $3.0$ & $51.40$ & $51.70$ & $0.30$ & $-0.58$ & $0.23$ & $-0.07$ \\
$-0.60$ & $-0.40$ & $4.0$ & $50.40$ & $50.40$ & $0.00$ & $-0.26$ & $0.10$ & $0.10$ \\
$-0.60$ & $-0.40$ & $5.0$ & $49.00$ & $49.10$ & $0.10$ & $-0.12$ & $0.05$ & $-0.05$ \\
\hline
$-0.40$ & $-0.30$ & $0.5$ & $51.70$ & $54.50$ & $2.80$ & $-8.93$ & $2.68$ & $-0.12$ \\
$-0.40$ & $-0.30$ & $1.0$ & $52.70$ & $54.40$ & $1.70$ & $-5.43$ & $1.63$ & $-0.07$ \\
$-0.40$ & $-0.30$ & $1.5$ & $52.55$ & $53.60$ & $1.05$ & $-3.18$ & $0.95$ & $-0.10$ \\
$-0.40$ & $-0.30$ & $2.0$ & $52.20$ & $52.80$ & $0.60$ & $-1.94$ & $0.58$ & $-0.02$ \\
$-0.40$ & $-0.30$ & $3.0$ & $51.40$ & $51.60$ & $0.20$ & $-0.81$ & $0.24$ & $0.04$ \\
$-0.40$ & $-0.30$ & $4.0$ & $50.30$ & $50.40$ & $0.10$ & $-0.36$ & $0.11$ & $0.01$ \\
$-0.40$ & $-0.30$ & $5.0$ & $48.90$ & $49.00$ & $0.10$ & $-0.16$ & $0.05$ & $-0.05$ \\
$-0.40$ & $0.10$ & $0.5$ & $51.70$ & $51.00$ & $-0.70$ & $-8.27$ & $-0.83$ & $-0.13$ \\
$-0.40$ & $0.10$ & $1.0$ & $52.70$ & $52.20$ & $-0.50$ & $-5.68$ & $-0.57$ & $-0.07$ \\
$-0.40$ & $0.10$ & $1.5$ & $52.55$ & $52.30$ & $-0.25$ & $-3.46$ & $-0.35$ & $-0.10$ \\
$-0.40$ & $0.10$ & $2.0$ & $52.20$ & $52.00$ & $-0.20$ & $-2.01$ & $-0.20$ & $0.00$ \\
$-0.40$ & $0.10$ & $3.0$ & $51.40$ & $51.30$ & $-0.10$ & $-0.63$ & $-0.06$ & $0.04$ \\
$-0.40$ & $0.10$ & $4.0$ & $50.30$ & $50.20$ & $-0.10$ & $-0.18$ & $-0.02$ & $0.08$ \\
$-0.40$ & $0.10$ & $5.0$ & $48.90$ & $48.90$ & $0.00$ & $-0.05$ & $-0.01$ & $-0.01$ \\
\hline
\end{tabular}
\end{center}
\end{table}

\newpage
\begin{table*}
{\bf Table \ref{tab:BetaCorrections0p10} continued}
\vspace{0.2cm}
\begin{center}
\begin{tabular}{|c|c|c|c|c|c|c|c|c|}
\hline
$\alpha$ & $\beta$ & $\lambda$ & $\tilde{\omega}_0$ & $\tilde{\omega}$ & $\tilde{\omega}-\tilde{\omega_0}$ & $c$ & $\omega-\omega_0$ & $\delta$ \\
\hline
$-0.20$ & $-0.30$ & $0.5$ & $51.70$ & $54.70$ & $3.00$ & $-9.86$ & $2.96$ & $-0.04$ \\
$-0.20$ & $-0.30$ & $1.0$ & $52.30$ & $54.30$ & $2.00$ & $-6.34$ & $1.90$ & $-0.10$ \\
$-0.20$ & $-0.30$ & $1.5$ & $52.15$ & $53.40$ & $1.25$ & $-3.88$ & $1.16$ & $-0.09$ \\
$-0.20$ & $-0.30$ & $2.0$ & $51.90$ & $52.70$ & $0.80$ & $-2.44$ & $0.73$ & $-0.07$ \\
$-0.20$ & $-0.30$ & $3.0$ & $51.10$ & $51.50$ & $0.40$ & $-1.05$ & $0.31$ & $-0.09$ \\
$-0.20$ & $-0.30$ & $4.0$ & $50.00$ & $50.20$ & $0.20$ & $-0.46$ & $0.14$ & $-0.06$ \\
$-0.20$ & $-0.30$ & $5.0$ & $48.80$ & $48.90$ & $0.10$ & $-0.20$ & $0.06$ & $-0.04$ \\
$-0.20$ & $0.10$ & $0.5$ & $51.70$ & $50.90$ & $-0.80$ & $-7.91$ & $-0.79$ & $0.01$ \\
$-0.20$ & $0.10$ & $1.0$ & $52.30$ & $51.80$ & $-0.50$ & $-5.93$ & $-0.59$ & $-0.09$ \\
$-0.20$ & $0.10$ & $1.5$ & $52.15$ & $51.90$ & $-0.25$ & $-3.77$ & $-0.38$ & $-0.13$ \\
$-0.20$ & $0.10$ & $2.0$ & $51.90$ & $51.70$ & $-0.20$ & $-2.24$ & $-0.22$ & $-0.02$ \\
$-0.20$ & $0.10$ & $3.0$ & $51.10$ & $51.00$ & $-0.10$ & $-0.72$ & $-0.07$ & $0.03$ \\
$-0.20$ & $0.10$ & $4.0$ & $50.00$ & $50.00$ & $0.00$ & $-0.21$ & $-0.02$ & $-0.02$ \\
$-0.20$ & $0.10$ & $5.0$ & $48.80$ & $48.70$ & $-0.10$ & $-0.06$ & $-0.01$ & $0.09$ \\
\hline
$0.20$ & $-0.20$ & $0.5$ & $52.50$ & $54.50$ & $2.00$ & $-11.21$ & $2.24$ & $0.24$ \\
$0.20$ & $-0.20$ & $1.0$ & $51.40$ & $53.00$ & $1.60$ & $-8.02$ & $1.60$ & $0.00$ \\
$0.20$ & $-0.20$ & $1.5$ & $51.05$ & $52.00$ & $0.95$ & $-5.30$ & $1.06$ & $0.11$ \\
$0.20$ & $-0.20$ & $2.0$ & $50.70$ & $51.40$ & $0.70$ & $-3.51$ & $0.70$ & $0.00$ \\
$0.20$ & $-0.20$ & $3.0$ & $50.00$ & $50.40$ & $0.40$ & $-1.56$ & $0.31$ & $-0.09$ \\
$0.20$ & $-0.20$ & $4.0$ & $49.20$ & $49.30$ & $0.10$ & $-0.69$ & $0.14$ & $0.04$ \\
$0.20$ & $-0.20$ & $5.0$ & $48.00$ & $48.10$ & $0.10$ & $-0.29$ & $0.06$ & $-0.04$ \\
$0.20$ & $0.20$ & $0.5$ & $52.50$ & $50.90$ & $-1.60$ & $-7.17$ & $-1.43$ & $0.17$ \\
$0.20$ & $0.20$ & $1.0$ & $51.40$ & $50.20$ & $-1.20$ & $-6.42$ & $-1.28$ & $-0.08$ \\
$0.20$ & $0.20$ & $1.5$ & $51.05$ & $50.10$ & $-0.95$ & $-4.39$ & $-0.88$ & $0.07$ \\
$0.20$ & $0.20$ & $2.0$ & $50.70$ & $50.10$ & $-0.60$ & $-2.72$ & $-0.54$ & $0.06$ \\
$0.20$ & $0.20$ & $3.0$ & $50.00$ & $49.70$ & $-0.30$ & $-0.90$ & $-0.18$ & $0.12$ \\
$0.20$ & $0.20$ & $4.0$ & $49.20$ & $49.00$ & $-0.20$ & $-0.27$ & $-0.05$ & $0.15$ \\
$0.20$ & $0.20$ & $5.0$ & $48.00$ & $48.00$ & $0.00$ & $-0.08$ & $-0.02$ & $-0.02$ \\
\hline
\end{tabular}
\end{center}
\end{table*}

\newpage
\begin{table*}
{\bf Table \ref{tab:BetaCorrections0p10} continued}
\vspace{0.2cm}
\begin{center}
\begin{tabular}{|c|c|c|c|c|c|c|c|c|}
\hline
$\alpha$ & $\beta$ & $\lambda$ & $\tilde{\omega}_0$ & $\tilde{\omega}$ & $\tilde{\omega}-\tilde{\omega_0}$ & $c$ & $\omega-\omega_0$ & $\delta$ \\
\hline
$0.40$ & $-0.10$ & $0.5$ & $53.50$ & $54.50$ & $1.00$ & $-11.70$ & $1.17$ & $0.17$ \\
$0.40$ & $-0.10$ & $1.0$ & $51.20$ & $52.00$ & $0.80$ & $-8.83$ & $0.88$ & $0.08$ \\
$0.40$ & $-0.10$ & $1.5$ & $50.25$ & $50.80$ & $0.55$ & $-6.03$ & $0.60$ & $0.05$ \\
$0.40$ & $-0.10$ & $2.0$ & $49.80$ & $50.20$ & $0.40$ & $-4.08$ & $0.41$ & $0.01$ \\
$0.40$ & $-0.10$ & $3.0$ & $49.10$ & $49.30$ & $0.20$ & $-1.84$ & $0.18$ & $-0.02$ \\
$0.40$ & $-0.10$ & $4.0$ & $48.40$ & $48.50$ & $0.10$ & $-0.81$ & $0.08$ & $-0.02$ \\
$0.40$ & $-0.10$ & $5.0$ & $47.40$ & $47.50$ & $0.10$ & $-0.34$ & $0.03$ & $-0.07$ \\
$0.40$ & $0.30$ & $0.5$ & $53.50$ & $51.40$ & $-2.10$ & $-6.81$ & $-2.04$ & $0.06$ \\
$0.40$ & $0.30$ & $1.0$ & $51.20$ & $49.20$ & $-2.00$ & $-6.67$ & $-2.00$ & $0.00$ \\
$0.40$ & $0.30$ & $1.5$ & $50.25$ & $48.80$ & $-1.45$ & $-4.70$ & $-1.41$ & $0.04$ \\
$0.40$ & $0.30$ & $2.0$ & $49.80$ & $48.80$ & $-1.00$ & $-2.96$ & $-0.89$ & $0.11$ \\
$0.40$ & $0.30$ & $3.0$ & $49.10$ & $48.60$ & $-0.50$ & $-0.99$ & $-0.30$ & $0.20$ \\
$0.40$ & $0.30$ & $4.0$ & $48.40$ & $48.10$ & $-0.30$ & $-0.30$ & $-0.09$ & $0.21$ \\
$0.40$ & $0.30$ & $5.0$ & $47.40$ & $47.20$ & $-0.20$ & $-0.08$ & $-0.03$ & $0.17$ \\
\hline
$0.60$ & $-0.10$ & $0.5$ & $55.40$ & $56.40$ & $1.00$ & $-12.10$ & $1.21$ & $0.21$ \\
$0.60$ & $-0.10$ & $1.0$ & $51.40$ & $52.30$ & $0.90$ & $-9.62$ & $0.96$ & $0.06$ \\
$0.60$ & $-0.10$ & $1.5$ & $49.65$ & $50.30$ & $0.65$ & $-6.78$ & $0.68$ & $0.03$ \\
$0.60$ & $-0.10$ & $2.0$ & $48.80$ & $49.20$ & $0.40$ & $-4.66$ & $0.47$ & $0.07$ \\
$0.60$ & $-0.10$ & $3.0$ & $47.90$ & $48.20$ & $0.30$ & $-2.13$ & $0.21$ & $-0.09$ \\
$0.60$ & $-0.10$ & $4.0$ & $47.30$ & $47.40$ & $0.10$ & $-0.93$ & $0.09$ & $-0.01$ \\
$0.60$ & $-0.10$ & $5.0$ & $46.50$ & $46.60$ & $0.10$ & $-0.39$ & $0.04$ & $-0.06$ \\
$0.60$ & $0.30$ & $0.5$ & $55.40$ & $53.50$ & $-1.90$ & $-6.44$ & $-1.93$ & $-0.03$ \\
$0.60$ & $0.30$ & $1.0$ & $51.40$ & $49.30$ & $-2.10$ & $-6.92$ & $-2.08$ & $0.02$ \\
$0.60$ & $0.30$ & $1.5$ & $49.65$ & $48.00$ & $-1.65$ & $-5.02$ & $-1.50$ & $0.15$ \\
$0.60$ & $0.30$ & $2.0$ & $48.80$ & $47.60$ & $-1.20$ & $-3.19$ & $-0.96$ & $0.24$ \\
$0.60$ & $0.30$ & $3.0$ & $47.90$ & $47.30$ & $-0.60$ & $-1.09$ & $-0.33$ & $0.27$ \\
$0.60$ & $0.30$ & $4.0$ & $47.30$ & $46.90$ & $-0.40$ & $-0.33$ & $-0.10$ & $0.30$ \\
$0.60$ & $0.30$ & $5.0$ & $46.50$ & $46.30$ & $-0.20$ & $-0.09$ & $-0.03$ & $0.17$ \\
\hline
\end{tabular}
\end{center}
\end{table*}

\newpage
\begin{table}[h!]
{\bf \caption{\label{tab:Omega0p25}}}The f\mbox{}itted values of the angle $\omega$ (in degrees) for $\lambda_0^{-1}=0.25$ and $\beta=0$.
\vspace{0.2cm}
\begin{center}
\begin{tabular}{|c|c|c|c|c|c|c|c|c|}
\hline
$\lambda \downarrow$, $\alpha \rightarrow$ & $-0.60$ & $-0.40$ & $-0.20$ & $-0.05$ & $0.05$ & $0.20$ & $0.40$ & $0.60$ \\
\hline
$0.1$ & $44.94$ & $46.03$ & $48.83$ & $51.47$ & $52.65$ & $54.63$ & $58.03$ & $61.45$ \\
$0.2$ & $48.34$ & $48.63$ & $49.92$ & $51.38$ & $52.30$ & $53.62$ & $56.10$ & $58.90$ \\
$0.3$ & $50.11$ & $49.98$ & $50.51$ & $51.28$ & $51.98$ & $52.78$ & $54.55$ & $56.82$ \\
$0.4$ & $51.11$ & $50.74$ & $50.84$ & $51.17$ & $51.68$ & $52.09$ & $53.31$ & $55.13$ \\
$0.5$ & $51.66$ & $51.16$ & $51.01$ & $51.06$ & $51.40$ & $51.52$ & $52.33$ & $53.77$ \\
$0.6$ & $51.93$ & $51.37$ & $51.07$ & $50.94$ & $51.13$ & $51.06$ & $51.55$ & $52.66$ \\
$0.7$ & $52.02$ & $51.45$ & $51.07$ & $50.81$ & $50.88$ & $50.68$ & $50.93$ & $51.75$ \\
$0.8$ & $51.99$ & $51.43$ & $51.01$ & $50.67$ & $50.63$ & $50.36$ & $50.42$ & $50.99$ \\
$0.9$ & $51.87$ & $51.35$ & $50.90$ & $50.52$ & $50.39$ & $50.08$ & $49.98$ & $50.34$ \\
$1.0$ & $51.69$ & $51.23$ & $50.76$ & $50.35$ & $50.16$ & $49.82$ & $49.60$ & $49.77$ \\
$1.1$ & $51.47$ & $51.07$ & $50.60$ & $50.16$ & $49.93$ & $49.57$ & $49.25$ & $49.25$ \\
$1.2$ & $51.22$ & $50.88$ & $50.41$ & $49.96$ & $49.69$ & $49.32$ & $48.91$ & $48.77$ \\
$1.3$ & $50.95$ & $50.67$ & $50.19$ & $49.74$ & $49.46$ & $49.05$ & $48.57$ & $48.33$ \\
$1.4$ & $50.67$ & $50.44$ & $49.96$ & $49.50$ & $49.21$ & $48.77$ & $48.23$ & $47.92$ \\
$1.5$ & $50.37$ & $50.19$ & $49.70$ & $49.23$ & $48.95$ & $48.47$ & $47.89$ & $47.53$ \\
$1.6$ & $50.07$ & $49.93$ & $49.43$ & $48.94$ & $48.69$ & $48.16$ & $47.54$ & $47.17$ \\
$1.7$ & $49.77$ & $49.65$ & $49.13$ & $48.63$ & $48.41$ & $47.84$ & $47.21$ & $46.84$ \\
$1.8$ & $49.46$ & $49.35$ & $48.82$ & $48.29$ & $48.11$ & $47.53$ & $46.87$ & $46.51$ \\
$1.9$ & $49.15$ & $49.04$ & $48.49$ & $47.92$ & $47.79$ & $47.24$ & $46.54$ & $46.18$ \\
$2.0$ & $48.84$ & $48.72$ & $48.14$ & $47.53$ & $47.45$ & $46.99$ & $46.20$ & $45.78$ \\
\hline
\end{tabular}
\end{center}
\end{table}

\newpage
\begin{table}[h!]
{\bf \caption{\label{tab:CorrectionForOmega0p25}}}The $\beta$-correction factors $c(\alpha, \lambda_0^{-1}=0.25; \lambda)$ (in degrees), 
def\mbox{}ined in Eq.~(\ref{eq:DefinitionOfc}); upper part: values for $\beta < 0$, lower part: values for $\beta > 0$.
\vspace{0.2cm}
\begin{center}
\begin{tabular}{|c|c|c|c|c|c|c|c|c|}
\hline
$\lambda \downarrow$, $\alpha \rightarrow$ & $-0.60$ & $-0.40$ & $-0.20$ & $-0.05$ & $0.05$ & $0.20$ & $0.40$ & $0.60$ \\
\hline
$0.10$ & $-4.11$ & $-2.87$ & $-1.05$ & $0.61$ & $1.84$ & $3.83$ & $6.75$ & $9.94$ \\
$0.20$ & $-7.75$ & $-7.54$ & $-6.72$ & $-5.81$ & $-5.09$ & $-3.86$ & $-1.97$ & $0.16$ \\
$0.30$ & $-8.29$ & $-8.50$ & $-8.17$ & $-7.66$ & $-7.23$ & $-6.44$ & $-5.18$ & $-3.71$ \\
$0.40$ & $-7.97$ & $-8.40$ & $-8.37$ & $-8.12$ & $-7.87$ & $-7.39$ & $-6.58$ & $-5.59$ \\
$0.50$ & $-7.34$ & $-7.89$ & $-8.05$ & $-7.99$ & $-7.88$ & $-7.62$ & $-7.14$ & $-6.52$ \\
$0.60$ & $-6.61$ & $-7.23$ & $-7.52$ & $-7.59$ & $-7.58$ & $-7.49$ & $-7.25$ & $-6.91$ \\
$0.70$ & $-5.87$ & $-6.52$ & $-6.90$ & $-7.06$ & $-7.12$ & $-7.15$ & $-7.11$ & $-6.97$ \\
$0.80$ & $-5.17$ & $-5.82$ & $-6.25$ & $-6.48$ & $-6.59$ & $-6.72$ & $-6.81$ & $-6.84$ \\
$0.90$ & $-4.52$ & $-5.15$ & $-5.62$ & $-5.89$ & $-6.04$ & $-6.23$ & $-6.44$ & $-6.59$ \\
$1.00$ & $-3.93$ & $-4.54$ & $-5.02$ & $-5.31$ & $-5.49$ & $-5.73$ & $-6.01$ & $-6.25$ \\
$1.10$ & $-3.39$ & $-3.98$ & $-4.46$ & $-4.77$ & $-4.96$ & $-5.24$ & $-5.57$ & $-5.88$ \\
$1.20$ & $-2.92$ & $-3.47$ & $-3.94$ & $-4.26$ & $-4.47$ & $-4.76$ & $-5.13$ & $-5.48$ \\
$1.30$ & $-2.50$ & $-3.02$ & $-3.47$ & $-3.80$ & $-4.00$ & $-4.30$ & $-4.69$ & $-5.08$ \\
$1.40$ & $-2.13$ & $-2.61$ & $-3.05$ & $-3.37$ & $-3.57$ & $-3.88$ & $-4.28$ & $-4.68$ \\
$1.50$ & $-1.81$ & $-2.26$ & $-2.67$ & $-2.98$ & $-3.18$ & $-3.48$ & $-3.88$ & $-4.29$ \\
$1.60$ & $-1.53$ & $-1.94$ & $-2.34$ & $-2.63$ & $-2.82$ & $-3.12$ & $-3.52$ & $-3.92$ \\
$1.70$ & $-1.29$ & $-1.67$ & $-2.04$ & $-2.31$ & $-2.50$ & $-2.79$ & $-3.17$ & $-3.57$ \\
$1.80$ & $-1.08$ & $-1.42$ & $-1.77$ & $-2.03$ & $-2.21$ & $-2.48$ & $-2.86$ & $-3.24$ \\
$1.90$ & $-0.90$ & $-1.21$ & $-1.53$ & $-1.78$ & $-1.95$ & $-2.21$ & $-2.56$ & $-2.93$ \\
$2.00$ & $-0.75$ & $-1.03$ & $-1.32$ & $-1.56$ & $-1.72$ & $-1.96$ & $-2.30$ & $-2.65$ \\
\hline
$0.10$ & $-1.02$ & $0.09$ & $2.13$ & $4.24$ & $5.91$ & $8.82$ & $13.43$ & $18.84$ \\
$0.20$ & $-4.38$ & $-4.91$ & $-4.57$ & $-3.78$ & $-3.00$ & $-1.47$ & $1.23$ & $4.66$ \\
$0.30$ & $-4.97$ & $-5.93$ & $-6.13$ & $-5.80$ & $-5.36$ & $-4.38$ & $-2.49$ & $0.04$ \\
$0.40$ & $-4.88$ & $-5.98$ & $-6.41$ & $-6.32$ & $-6.06$ & $-5.41$ & $-4.04$ & $-2.10$ \\
$0.50$ & $-4.58$ & $-5.71$ & $-6.25$ & $-6.31$ & $-6.18$ & $-5.76$ & $-4.76$ & $-3.29$ \\
$0.60$ & $-4.22$ & $-5.32$ & $-5.92$ & $-6.08$ & $-6.04$ & $-5.79$ & $-5.09$ & $-3.99$ \\
$0.70$ & $-3.83$ & $-4.88$ & $-5.52$ & $-5.74$ & $-5.77$ & $-5.65$ & $-5.18$ & $-4.37$ \\
\hline
\end{tabular}
\end{center}
\end{table}

\newpage
\begin{table*}
{\bf Table \ref{tab:CorrectionForOmega0p25} continued}
\vspace{0.2cm}
\begin{center}
\begin{tabular}{|c|c|c|c|c|c|c|c|c|}
\hline
$\lambda \downarrow$, $\alpha \rightarrow$ & $-0.60$ & $-0.40$ & $-0.20$ & $-0.05$ & $0.05$ & $0.20$ & $0.40$ & $0.60$ \\
\hline
$0.80$ & $-3.45$ & $-4.44$ & $-5.08$ & $-5.34$ & $-5.42$ & $-5.40$ & $-5.11$ & $-4.54$ \\
$0.90$ & $-3.08$ & $-4.00$ & $-4.62$ & $-4.91$ & $-5.02$ & $-5.07$ & $-4.92$ & $-4.54$ \\
$1.00$ & $-2.73$ & $-3.58$ & $-4.17$ & $-4.47$ & $-4.60$ & $-4.70$ & $-4.65$ & $-4.41$ \\
$1.10$ & $-2.41$ & $-3.18$ & $-3.73$ & $-4.03$ & $-4.17$ & $-4.30$ & $-4.33$ & $-4.20$ \\
$1.20$ & $-2.12$ & $-2.80$ & $-3.32$ & $-3.60$ & $-3.75$ & $-3.90$ & $-3.98$ & $-3.92$ \\
$1.30$ & $-1.84$ & $-2.46$ & $-2.93$ & $-3.20$ & $-3.34$ & $-3.50$ & $-3.61$ & $-3.61$ \\
$1.40$ & $-1.60$ & $-2.14$ & $-2.57$ & $-2.81$ & $-2.95$ & $-3.11$ & $-3.24$ & $-3.28$ \\
$1.50$ & $-1.38$ & $-1.86$ & $-2.23$ & $-2.46$ & $-2.59$ & $-2.74$ & $-2.88$ & $-2.94$ \\
$1.60$ & $-1.19$ & $-1.60$ & $-1.94$ & $-2.14$ & $-2.26$ & $-2.40$ & $-2.54$ & $-2.62$ \\
$1.70$ & $-1.01$ & $-1.37$ & $-1.67$ & $-1.85$ & $-1.96$ & $-2.09$ & $-2.22$ & $-2.31$ \\
$1.80$ & $-0.86$ & $-1.17$ & $-1.43$ & $-1.59$ & $-1.69$ & $-1.81$ & $-1.93$ & $-2.02$ \\
$1.90$ & $-0.73$ & $-1.00$ & $-1.22$ & $-1.36$ & $-1.45$ & $-1.55$ & $-1.67$ & $-1.75$ \\
$2.00$ & $-0.62$ & $-0.85$ & $-1.04$ & $-1.16$ & $-1.23$ & $-1.33$ & $-1.44$ & $-1.51$ \\
\hline
\end{tabular}
\end{center}
\end{table*}

\newpage
\begin{table}[h!]
{\bf \caption{\label{tab:Omega0p50}}}The values of the angle $\omega$ (in degrees) for $\lambda_0^{-1}=0.50$ and $\beta=0$.
\vspace{0.2cm}
\begin{center}
\begin{tabular}{|c|c|c|c|c|c|c|c|c|}
\hline
$\lambda \downarrow$, $\alpha \rightarrow$ & $-0.60$ & $-0.40$ & $-0.20$ & $-0.05$ & $0.05$ & $0.20$ & $0.40$ & $0.60$ \\
\hline
$0.1$ & $44.52$ & $46.04$ & $48.23$ & $50.10$ & $51.31$ & $53.10$ & $56.01$ & $58.89$ \\
$0.2$ & $47.83$ & $48.09$ & $48.92$ & $50.01$ & $50.77$ & $52.00$ & $53.86$ & $56.17$ \\
$0.3$ & $49.30$ & $48.99$ & $49.18$ & $49.80$ & $50.24$ & $51.04$ & $52.31$ & $54.22$ \\
$0.4$ & $49.94$ & $49.36$ & $49.20$ & $49.49$ & $49.69$ & $50.17$ & $51.10$ & $52.62$ \\
$0.5$ & $50.10$ & $49.41$ & $49.05$ & $49.09$ & $49.14$ & $49.38$ & $50.06$ & $51.19$ \\
$0.6$ & $49.98$ & $49.26$ & $48.79$ & $48.63$ & $48.58$ & $48.63$ & $49.08$ & $49.88$ \\
$0.7$ & $49.68$ & $48.96$ & $48.44$ & $48.12$ & $48.00$ & $47.92$ & $48.12$ & $48.68$ \\
$0.8$ & $49.26$ & $48.54$ & $48.01$ & $47.58$ & $47.41$ & $47.24$ & $47.22$ & $47.59$ \\
$0.9$ & $48.75$ & $48.02$ & $47.53$ & $47.03$ & $46.81$ & $46.56$ & $46.40$ & $46.61$ \\
$1.0$ & $48.19$ & $47.41$ & $46.99$ & $46.48$ & $46.19$ & $45.89$ & $45.69$ & $45.80$ \\
\hline
\end{tabular}
\end{center}
\end{table}

\newpage
\begin{table}[h!]
{\bf \caption{\label{tab:CorrectionForOmega0p50}}}The $\beta$-correction factors $c(\alpha, \lambda_0^{-1}=0.50; \lambda)$ (in degrees), 
def\mbox{}ined in Eq.~(\ref{eq:DefinitionOfc}); upper part: values for $\beta < 0$, lower part: values for $\beta > 0$.
\vspace{0.2cm}
\begin{center}
\begin{tabular}{|c|c|c|c|c|c|c|c|c|}
\hline
$\lambda \downarrow$, $\alpha \rightarrow$ & $-0.60$ & $-0.40$ & $-0.20$ & $-0.05$ & $0.05$ & $0.20$ & $0.40$ & $0.60$ \\
\hline
$0.10$ & $-1.43$ & $-0.31$ & $1.43$ & $3.08$ & $4.32$ & $6.39$ & $9.51$ & $13.02$ \\
$0.20$ & $-5.28$ & $-5.26$ & $-4.50$ & $-3.53$ & $-2.73$ & $-1.29$ & $1.02$ & $3.74$ \\
$0.30$ & $-6.06$ & $-6.45$ & $-6.16$ & $-5.57$ & $-5.03$ & $-4.01$ & $-2.30$ & $-0.23$ \\
$0.40$ & $-5.88$ & $-6.47$ & $-6.46$ & $-6.15$ & $-5.82$ & $-5.14$ & $-3.94$ & $-2.43$ \\
$0.50$ & $-5.30$ & $-5.95$ & $-6.12$ & $-6.00$ & $-5.81$ & $-5.40$ & $-4.60$ & $-3.56$ \\
$0.60$ & $-4.55$ & $-5.19$ & $-5.45$ & $-5.45$ & $-5.37$ & $-5.13$ & $-4.64$ & $-3.95$ \\
$0.70$ & $-3.77$ & $-4.36$ & $-4.65$ & $-4.72$ & $-4.70$ & $-4.59$ & $-4.29$ & $-3.86$ \\
$0.80$ & $-3.04$ & $-3.56$ & $-3.84$ & $-3.94$ & $-3.96$ & $-3.92$ & $-3.76$ & $-3.50$ \\
$0.90$ & $-2.41$ & $-2.83$ & $-3.09$ & $-3.20$ & $-3.23$ & $-3.24$ & $-3.17$ & $-3.02$ \\
$1.00$ & $-1.87$ & $-2.22$ & $-2.44$ & $-2.54$ & $-2.58$ & $-2.61$ & $-2.59$ & $-2.51$ \\
\hline
$0.10$ & $1.86$ & $3.34$ & $5.11$ & $6.61$ & $7.68$ & $9.41$ & $11.90$ & $14.59$ \\
$0.20$ & $-2.38$ & $-2.02$ & $-1.22$ & $-0.37$ & $0.29$ & $1.43$ & $3.18$ & $5.19$ \\
$0.30$ & $-3.64$ & $-3.74$ & $-3.40$ & $-2.91$ & $-2.48$ & $-1.71$ & $-0.46$ & $1.03$ \\
$0.40$ & $-3.90$ & $-4.23$ & $-4.17$ & $-3.92$ & $-3.67$ & $-3.18$ & $-2.34$ & $-1.29$ \\
$0.50$ & $-3.68$ & $-4.12$ & $-4.23$ & $-4.14$ & $-4.01$ & $-3.73$ & $-3.19$ & $-2.50$ \\
$0.60$ & $-3.25$ & $-3.70$ & $-3.90$ & $-3.91$ & $-3.86$ & $-3.72$ & $-3.40$ & $-2.96$ \\
$0.70$ & $-2.74$ & $-3.16$ & $-3.39$ & $-3.45$ & $-3.45$ & $-3.40$ & $-3.23$ & $-2.97$ \\
$0.80$ & $-2.23$ & $-2.60$ & $-2.82$ & $-2.91$ & $-2.93$ & $-2.93$ & $-2.86$ & $-2.71$ \\
$0.90$ & $-1.76$ & $-2.08$ & $-2.28$ & $-2.37$ & $-2.41$ & $-2.43$ & $-2.41$ & $-2.33$ \\
$1.00$ & $-1.37$ & $-1.63$ & $-1.80$ & $-1.88$ & $-1.92$ & $-1.95$ & $-1.96$ & $-1.93$ \\
\hline
\end{tabular}
\end{center}
\end{table}

\newpage
\begin{table}[h!]
{\bf \caption{\label{tab:Omega1p00}}}The values of the angle $\omega$ (in degrees) for $\lambda_0^{-1}=1.00$ and $\beta=0$.
\vspace{0.2cm}
\begin{center}
\begin{tabular}{|c|c|c|c|c|c|c|c|c|}
\hline
$\lambda \downarrow$, $\alpha \rightarrow$ & $-0.60$ & $-0.40$ & $-0.20$ & $-0.05$ & $0.05$ & $0.20$ & $0.40$ & $0.60$ \\
\hline
$0.10$ & $42.89$ & $44.50$ & $46.34$ & $47.88$ & $48.91$ & $50.45$ & $52.63$ & $54.86$ \\
$0.15$ & $44.68$ & $45.37$ & $46.48$ & $47.59$ & $48.37$ & $49.69$ & $51.39$ & $53.27$ \\
$0.20$ & $45.74$ & $45.89$ & $46.53$ & $47.28$ & $47.85$ & $48.91$ & $50.26$ & $51.83$ \\
$0.25$ & $46.34$ & $46.17$ & $46.48$ & $46.96$ & $47.35$ & $48.14$ & $49.22$ & $50.51$ \\
$0.30$ & $46.65$ & $46.25$ & $46.33$ & $46.59$ & $46.84$ & $47.37$ & $48.24$ & $49.30$ \\ 
$0.35$ & $46.72$ & $46.19$ & $46.08$ & $46.19$ & $46.32$ & $46.63$ & $47.33$ & $48.17$ \\
$0.40$ & $46.63$ & $46.00$ & $45.75$ & $45.72$ & $45.77$ & $45.93$ & $46.44$ & $47.11$ \\
$0.45$ & $46.41$ & $45.71$ & $45.33$ & $45.19$ & $45.20$ & $45.27$ & $45.58$ & $46.10$ \\
$0.50$ & $46.08$ & $45.32$ & $44.82$ & $44.58$ & $44.58$ & $44.66$ & $44.72$ & $45.14$ \\
\hline
\end{tabular}
\end{center}
\end{table}

\newpage
\begin{table}[h!]
{\bf \caption{\label{tab:CorrectionForOmega1p00}}}The $\beta$-correction factors $c(\alpha, \lambda_0^{-1}=1.00; \lambda)$ (in degrees), 
def\mbox{}ined in Eq.~(\ref{eq:DefinitionOfc}); upper part: values for $\beta < 0$, lower part: values for $\beta > 0$.
\vspace{0.2cm}
\begin{center}
\begin{tabular}{|c|c|c|c|c|c|c|c|c|}
\hline
$\lambda \downarrow$, $\alpha \rightarrow$ & $-0.60$ & $-0.40$ & $-0.20$ & $-0.05$ & $0.05$ & $0.20$ & $0.40$ & $0.60$ \\
\hline
$0.10$ & $-0.48$ & $0.82$ & $2.13$ & $3.11$ & $3.76$ & $4.74$ & $6.04$ & $7.34$ \\
$0.15$ & $-3.19$ & $-2.30$ & $-1.41$ & $-0.74$ & $-0.29$ & $0.38$ & $1.27$ & $2.17$ \\
$0.20$ & $-4.12$ & $-3.51$ & $-2.90$ & $-2.44$ & $-2.13$ & $-1.68$ & $-1.07$ & $-0.46$ \\
$0.25$ & $-4.24$ & $-3.82$ & $-3.41$ & $-3.10$ & $-2.89$ & $-2.58$ & $-2.16$ & $-1.75$ \\
$0.30$ & $-3.98$ & $-3.70$ & $-3.42$ & $-3.21$ & $-3.07$ & $-2.86$ & $-2.58$ & $-2.30$ \\
$0.35$ & $-3.56$ & $-3.38$ & $-3.19$ & $-3.05$ & $-2.95$ & $-2.81$ & $-2.62$ & $-2.43$ \\
$0.40$ & $-3.09$ & $-2.97$ & $-2.84$ & $-2.74$ & $-2.68$ & $-2.59$ & $-2.46$ & $-2.33$ \\
$0.45$ & $-2.62$ & $-2.54$ & $-2.46$ & $-2.39$ & $-2.35$ & $-2.29$ & $-2.20$ & $-2.12$ \\
$0.50$ & $-2.19$ & $-2.14$ & $-2.08$ & $-2.04$ & $-2.01$ & $-1.97$ & $-1.92$ & $-1.86$ \\
\hline
$0.10$ & $3.19$ & $4.48$ & $5.83$ & $6.89$ & $7.61$ & $8.74$ & $10.30$ & $11.94$ \\
$0.15$ & $0.10$ & $0.79$ & $1.70$ & $2.51$ & $3.10$ & $4.06$ & $5.48$ & $7.04$ \\
$0.20$ & $-1.42$ & $-1.10$ & $-0.51$ & $0.09$ & $0.56$ & $1.34$ & $2.54$ & $3.90$ \\
$0.25$ & $-2.03$ & $-1.94$ & $-1.55$ & $-1.12$ & $-0.76$ & $-0.15$ & $0.82$ & $1.94$ \\
$0.30$ & $-2.15$ & $-2.17$ & $-1.94$ & $-1.63$ & $-1.36$ & $-0.89$ & $-0.14$ & $0.75$ \\
$0.35$ & $-2.01$ & $-2.09$ & $-1.95$ & $-1.74$ & $-1.54$ & $-1.19$ & $-0.61$ & $0.08$ \\
$0.40$ & $-1.76$ & $-1.86$ & $-1.79$ & $-1.64$ & $-1.50$ & $-1.24$ & $-0.80$ & $-0.28$ \\
$0.45$ & $-1.48$ & $-1.58$ & $-1.55$ & $-1.44$ & $-1.35$ & $-1.16$ & $-0.83$ & $-0.44$ \\
$0.50$ & $-1.20$ & $-1.30$ & $-1.29$ & $-1.22$ & $-1.15$ & $-1.01$ & $-0.78$ & $-0.49$ \\
\hline
\end{tabular}
\end{center}
\end{table}

\clearpage
% ============= FIGURE 1

\clearpage
% ============= FIGURE 2

\clearpage
% ============= FIGURE 3

\clearpage
% ============= FIGURE 4

\clearpage
% ============= FIGURE 5

\clearpage
% ============= FIGURE 6

\clearpage
% ============= FIGURE 7

\clearpage
% ============= FIGURE 8

\clearpage
% ============= FIGURE 9

\clearpage
% ============= FIGURE 10

\clearpage
% ============= FIGURE 11

\clearpage
% ============= FIGURE 12

\clearpage
% ============= FIGURE 13

\clearpage
\appendix \section{\label{sec:Misprints}Misprints in Refs.~\cite{SH} and \cite{HS}}

Some misprints have been found in Refs.~\cite{SH} and \cite{HS}.

To start with Appendix B of Ref.~\cite{SH}, the function $B(\xi)$ seems to bear the dimensions of stress (Pa, in SI), e.g., see expression (B$7$); given the 
form of expression (B$12$), the same goes to the coef\mbox{}f\mbox{}icients $a_k$. If one now considers expression (B$15$), the stress intensity factor $K$ should also 
have the same dimensions; however, stress intensity factors bear the dimension Pa$\sqrt{\rm m}$ (in SI). In retrospect, there is a missing factor somewhere 
(a square root of a length). Given, however, that a ratio of the real and imaginary parts of $K$ is to be taken in order to extract the value of $\omega$, any 
common factor will f\mbox{}inally drop out.

Concerning Appendix C, the expressions for $Y_1$ and $Y_2$, appearing on page $1353$, are not correct; a factor $h$ has been omitted. The 
correct expressions read as
\[
Y_1=-\Bigg\{ \frac{2+\Lambda+\Pi}{2} + [(1+\Pi)(h+d)+(\Lambda-\Pi)h] \lambda \Bigg\}e^{-(h+d) \lambda}
\]
and
\[
Y_2=+\Bigg\{ \frac{\Lambda-\Pi}{2} - [(1+\Pi)(h+d)+(\Lambda-\Pi)h] \lambda \Bigg\}e^{-(h+d) \lambda} \, .
\]

A factor of $2$ has been omitted in the $\Sigma \xi$ and $\Sigma \xi_1$ terms in the two numerators in expressions ($3.27$) of Ref.~\cite{HS}. The expressions 
(A$1$) of Ref.~\cite{SH} are correct. (Take also into account the change of notation between Refs.~\cite{SH} and \cite{HS}; $\lambda$ in the former is denoted 
as $\xi$ in the latter.)

\end{document}